\documentclass[aip,jcp,amsmath,amssymb,reprint]{revtex4-2}

\pdfoutput=1
\usepackage{epsfig}
\usepackage{amsmath}
\usepackage{booktabs}
\usepackage{multirow}
\usepackage{graphicx}
\usepackage{xcolor}
\usepackage{mathtools}
\usepackage{amssymb}
\usepackage{ragged2e}
\usepackage{threeparttable,booktabs}
\usepackage{diagbox}
\usepackage{float}
\usepackage[section]{placeins}
\usepackage{hhline}
\usepackage{bm}
\usepackage{dsfont}
\usepackage{siunitx,booktabs}
\usepackage[referable]{threeparttablex}
\usepackage[breaklinks=true,colorlinks=true,linkcolor=blue,urlcolor=blue,citecolor=blue,bookmarks=true,bookmarksopenlevel=2]{hyperref}
\DeclareSIUnit\angstrom{\text {Å}}

\AtBeginDocument{%
  \cmidrulewidth=.03em
  \cmidrulesep=\doublerulesep
  \cmidrulekern=.5em
  \defaultaddspace=.5em
}

\newcommand{\RUG}{
Van Swinderen Institute for Particle Physics and Gravity,
University of Groningen, Nijenborgh 4, 9747 AG Groningen, The Netherlands\\
\vspace*{0.3cm}
}

\newcommand{\UW}{
Faculty of Physics, University of Warsaw, Pasteura 5, 02-093 Warsaw, Poland\\
\vspace*{0.3cm}
}

\newcommand{\CU}{
Department of Physical and Theoretical Chemistry, Faculty of Natural Sciences, Comenius University, Mlynská dolina, 84215 Bratislava, Slovakia\\
\vspace*{0.3cm}
}

\newcommand{\IMIT}{
Instituto de Modelado e Innovación Tecnológica (UNNE-CONICET),
Facultad de Ciencias Exactas y Naturales y Agrimensura,
Universidad Nacional del Nordeste, Av. Libertad 5460, Corrientes, Argentina\\
\vspace*{0.3cm}
}

\newcommand{\Nikhef}{
Nikhef, National Institute for Subatomic Physics, Amsterdam, The Netherlands\\
\vspace*{0.3cm}
}

\begin{document}

\title{$\mathcal{P,T}$-odd effects in YbCu, YbAg and YbAu}

\author{Johan David Polet}
\affiliation{\RUG}

\author{Yuly Chamorro}
\affiliation{\RUG}
\affiliation{\Nikhef}

\author{Lukáš F. Pašteka}
\affiliation{\RUG}
\affiliation{\Nikhef}
\affiliation{\CU}

\author{Steven Hoekstra}
\affiliation{\RUG}
\affiliation{\Nikhef}

\author{Michał Tomza}
\affiliation{\UW}

\author{Anastasia Borschevsky}
\affiliation{\RUG}
\affiliation{\Nikhef}

\author{I.~Agustín Aucar}
\email[Author to whom correspondence should be addressed. Electronic mail: ]{agustin.aucar@conicet.gov.ar}
\affiliation{\RUG}
\affiliation{\Nikhef}
\affiliation{\IMIT}


\date{\today}

\begin{abstract}
In this work, the molecular enhancement factors of the $\mathcal{P,T}$-odd interactions involving the electron electric dipole moment ($W_\mathrm{d}$) and the scalar-pseudoscalar nucleon-electron couplings ($W_\mathrm{s}$) are computed for the ground state of the bimetallic molecules YbCu, YbAg and YbAu. These systems offer a promising venue for creating cold molecules by associating laser cooled atoms.
The relativistic coupled-cluster approach is used in the calculations and a thorough uncertainty analysis is performed to give accurate and reliable uncertainties to the obtained values. 
Furthermore, an in-depth investigation of the different electronic structure effects that determine the magnitude of the calculated enhancement factors is carried out, and two different schemes for computing $W_\mathrm{d}$ are compared.
The recommended values for the enhancement factors are $(13.24\pm0.03)\times10^{24}\frac{h\,\text{Hz}}{e\,\text{cm}}$, $(12.15\pm0.08)\times10^{24}\frac{h\,\text{Hz}}{e\,\text{cm}}$ and $(2.13\pm0.28)\times10^{24}\frac{h\,\text{Hz}}{e\,\text{cm}}$ for $W_\mathrm{d}$, and $(-48.36\pm0.18)\;h\,\text{kHz}$, $(-45.51\pm0.43)\;h\,\text{kHz}$ and $(5.31\pm1.80)\;h\,\text{kHz}$ for $W_\mathrm{s}$, for YbCu, YbAg and YbAu, respectively.
\end{abstract}

\maketitle

\section{Introduction}\label{sec:intro}

The current best description of elementary particles and their interactions is given by the Standard Model (SM) of particle physics~\cite{gaillard_1999_the}. This model is capable of explaining almost all experimental observations and of accurately predicting a wide range of diverse phenomena, which is why over time it has been consolidated as a well-tested physical theory. However, it does not address some important observed effects, such as the matter-antimatter asymmetry in the universe, the neutrino oscillations, and the existence and nature of dark matter and dark energy~\cite{safronova_2018_search,virdee_2016_beyond}. Over the past decades, many new theories and extensions of the SM have been proposed to explain these phenomena~\cite{virdee_2016_beyond, gouttenoire_2023_beyond}. Testing and restricting these theories is important for advancing our understanding of the fundamental laws of physics.

A promising way to test some of these theories is to search for effects due to the simultaneous non-conservation of spatial ($\mathcal{P}$) and time-reversal ($\mathcal{T}$) parities in atoms and molecules, such as those arising from the electric dipole moments (EDM) of electrons~\cite{ginges_2004_violations}. Interactions between these EDMs and electromagnetic fields violate both temporal and spatial invariance~\cite{sandars_1966_enhancement}. Within a $\mathcal{C}\mathcal{P}\mathcal{T}$-invariant theory ($\mathcal{C}$ refers to charge-conjugation symmetry), if $\mathcal{T}$ symmetry is violated, then the combined $\mathcal{CP}$ symmetry must also be non conserved, so that clearly $\mathcal{P,T}$ violation implies $\mathcal{CP}$ non conservation~\cite{safronova_2018_search}.

The sources of $\mathcal{CP}$ violation described by the SM lead to the prediction of a free-electron EDM $d_\mathrm{e}$ of approximately $5.8\times10^{-40}\,e\,\text{cm}$~\cite{Yamaguchi2020}. However, more sources of $\mathcal{CP}$ violation beyond those predicted by the SM are needed to explain, for example, the observed matter-antimatter asymmetry. The additional sources of $\mathcal{CP}$ violation would in turn lead to an increase in the magnitude of the electron EDM (eEDM), even bringing it into the reach of present day precision experiments~\cite{safronova_2018_search}. Experimental searches for these phenomena are currently being carried out in atoms and molecules, taking advantage of the enhancement of the atomic and molecular EDMs. These arise from $\mathcal{P,T}$-violating interactions, mainly those taking place between the eEDMs and the large internal atomic or molecular electric fields, and also other $\mathcal{CP}$-odd nucleon-electron and nucleon-nucleon interactions~\cite{sandars_1965_the,Kozlov85, sushkov_1978_parity, safronova_2018_search}. Currently, the lowest upper limit of the eEDM is set at $|d_\mathrm{e}| < 2.1\times10^{-29}\,e\,\text{cm}$. This upper limit was reported after measurements conducted by the National Institute of Science and Technology (NIST) on the HfF$^+$ molecular ion~\cite{RouCalWri23}, and combining these results with those obtained by the ACME Collaboration in their analysis of the ThO molecule~\cite{ACME2018}. It is important to stress that if $\mathcal{CP}$ violation is assumed to arise exclusively from $d_\mathrm{e}$ in the NIST experiment (i.e., if the scalar-pseudoscalar nucleon-electron couplings are neglected), then the upper limit would be $|d_\mathrm{e}| < 4.1\times10^{-30}\,e\,\text{cm}$. This lowest upper limit has already put considerable constraints on some of the theories beyond the SM~\cite{schwarzschild_2014_surprising,li_2024_does}.

In the ground state of a molecule with zero nuclear spins and a single unpaired electron there are two main contributions to its energy arising from $\mathcal{P,T}$-violating interactions. These arise from the interactions between the EDMs of the electrons and the electromagnetic fields, and from the $\mathcal{P,T}$-odd scalar-pseudoscalar nucleon-electron (S-PS-ne) neutral-current interactions~\cite{Gorshkov79,Kozlov85}. Since the effects of the eEDM and the S-PS-ne interactions are enhanced by the molecular electronic structure, the two corresponding molecular enhancement factors $W_\mathrm{d}$ (related to $d_\mathrm{e}$) and $W_\mathrm{s}$ (which enhances the S-PS-ne interactions) are of particular interest.

The choice of molecule for the measurements has a significant impact on the sensitivity due to, among others, the system-dependent enhancement of the $\mathcal{P,T}$-odd effects. Since in paramagnetic molecules containing only one heavy element this enhancement scales roughly as the cube of the atomic number of the heavier nucleus to which the unpaired electron is strongly linked to~\cite{gaul_2019_systematic,commins_2007_the}, some heavy-element-containing molecules have an advantage over other systems. Furthermore, practical experimental considerations play a crucial role in selecting a candidate for experiments. For example, the use of ultracold molecules increases interaction times and hence the experimental precision~\cite{CarDeMKre09}. Therefore, laser-coolability of the selected molecule provides a clear advantage. Various molecular properties relevant for precision measurements  (e.g., laser coolability and sensitivity to the measured phenomena, but also many others) can be determined theoretically before experimental investigations in support of such experiments. In particular, the enhancement factors cannot be measured and must be provided based on accurate electronic structure calculations.

In this work, we investigate the sensitivity of the YbCu, YbAg and YbAu paramagnetic molecules to $\mathcal{P,T}$-violating phenomena. These systems are of particular interest, since they contain two metal atoms. Theoretically, this means that these atoms can be laser-cooled separately and then associated into a molecule afterwards, eliminating the need to laser-cool the molecule as a whole~\cite{lee_2017_ultracold,verma_2020_electron}. So far, laser-cooling of Yb and Ag atoms has been demonstrated~\cite{uhlenberg_2000_magnetooptical,honda_1999_magnetooptical}. Laser-cooling of Cu and Au has not been demonstrated yet, but cooling schemes have been proposed~\cite{dzuba_2021_time}. Furthermore, these polar molecules have large molecular-frame electric dipole moments, due to the large electronegativity of the coinage-metal atoms, allowing for their easy polarization. In particular, YbAg is considered as a promising candidate for a next generation clock-transition eEDM measurement~\cite{verma_2020_electron}.

No experiments have been performed so far on YbCu, YbAg or YbAu, but several experimental groups pursue ultracold formation of other Ag-containing molecules~\cite{SmialkowskiPRA21,fleig_2021_theoretical,KlosNJP22,AurelienPRA23}. However, high-accuracy calculations of the potential energy curves, molecular-frame electric dipole moments, electric quadrupole moments, and static electric dipole polarizabilities of the present systems were recently performed~\cite{Tomza_2021}. Here we employ the four-component (4C) relativistic coupled-cluster (CC) approach to calculate the enhancement factors of the $\mathcal{P,T}$-violating interactions between the eEDMs and the electric fields in the systems, $W_\mathrm{d}$, and of the $\mathcal{P,T}$-odd S-PS-ne interactions, $W_\mathrm{s}$. The enhancement factors are determined for the ground states, $X \, ^2\Sigma^+_{\frac{1}{2}}$, of the three molecules. We also carry out an extensive computational study to assign uncertainties on the calculated values.

In paramagnetic systems containing at least one non-zero nuclear spin $I$, some internal nuclear interactions lead to nuclear spin-dependent molecular $\mathcal{P,T}$-violating effects, such as magnetic interactions between the electrons and the nuclear magnetic quadrupole moment (NMQM), which appears for nuclear spin $I \geq 1$~\cite{SusFlaKhr84,flambaum1994spin}. These effects are, however, outside the scope of this work.

Section~\ref{sec:Theory} covers the main theoretical aspects of this work, detailing how both enhancement factors can be obtained from effective Hamiltonians. Then, Section~\ref{sec:comp-det} contains a description of the methods employed to calculate these factors, as well as the scheme use for geometry optimization. Next, Section~\ref{sec:results} presents the obtained enhancement factors and their dependence on effects such as the choice of the nuclear charge density model, the method for treatment of electron correlation, the choice of the basis set, and the internuclear distances. Finally, Section~\ref{sec:conclusion} contains a concise summary of our findings.

\section{Theory}\label{sec:Theory}

The $\mathcal{P,T}$-violating interactions involving the EDMs of atoms or molecules produce non-zero linear Stark shifts in the limit of vanishingly small applied electric fields. These interactions originate from many different sources, but mainly from the EDMs of electrons and nucleons, as well as from the $\mathcal{P,T}$-violating nucleon-nucleon current interactions and the $\mathcal{P,T}$-odd electron–quark interactions~\cite{Engel2013}. In particular, for paramagnetic linear molecules in $X\,^2\Sigma_\frac{1}{2}$ ground states, such as the systems treated in this paper, the interactions between eEDMs and electromagnetic fields and the $\mathcal{P,T}$-odd S-PS-ne neutral-current interactions dominate~\cite{chupp_2015_electric}. In these cases, the effective spin-rotation Hamiltonian that includes only nuclear-spin-independent $\mathcal{P,T}$-odd interactions (i.e., neglecting the interactions between electrons and NMQMs) can be written as~\cite{Kozlov85,KozLab95}
\begin{equation}\label{eq:HeffSR}
\hat{H}_\text{sr}^{\mathcal{P,T}\mathrm{-odd}} = \left( \sum_K W_{\mathrm{s},K} \, k_{\mathrm{s},K} + W_\mathrm{d} \, d_\mathrm{e} \right) \hat{\Omega},
\end{equation}
where the $\mathcal{P,T}$-odd dimensionless operator $\hat{\Omega} = \hbar^{-1} \mathbf{J}_\mathrm{e} \cdot \mathbf{n}$ is the projection of the reduced total electronic angular momentum operator $\mathbf{J}_\mathrm{e}/\hbar$ along the direction of the molecular-frame electric dipole moment, which is given by the unit vector $\mathbf{n}$ (for a linear molecule, it points from the negatively charged region of the system to the positive one, along the internuclear axis). Here, $\hbar=h/(2\pi)$ is the reduced Planck constant. The sum in Eq.~\eqref{eq:HeffSR} runs over all the nuclei $K$ in the system. In systems where one nucleus is significantly heavier that the other (and where the unpaired electron is mostly located on that heavy nucleus), this sum typically reduces to a single term. We also note that $k_{\mathrm{s},K}$ is specific to each nucleus $K$, i.e., it depends on both the proton and the neutron numbers. Both these points will become important later and are discussed in Section~\ref{sec:wstheory}.

To compute the enhancement factors, the effects of the eEDM and the S-PS-ne interactions are taken as perturbations on the 4C relativistic Dirac--Coulomb (DC) Hamiltonian. Since $d_{\mathrm{e}}$ and $k_{\mathrm{s},K}$ are small quantities, the effects arising from both interactions are minute. Therefore, first-order perturbation treatment will already yield highly accurate results.

Within the Born-Oppenheimer approximation, the 4C DC (clamped-nuclei) Hamiltonian is given by
\begin{eqnarray}\label{eq:hamilelec}
\hat{H}^{(0)}&=&\sum_i\left[c\bm{\alpha}_i\cdot\hat{\bm{p}}_i+\beta_im_ec^2 - \sum_{K} e V_K(\bm{r}_i)\right]\nonumber \\
    &&
- \frac{1}{2} \sum_{i \neq j} e V_j(\bm{r}_i)
+ \frac{1}{2} \sum_{K \neq L} e Z_K V_L(\bm{R}_K),
\end{eqnarray}
\noindent where
\begin{eqnarray}\label{eq:Internal-potentials}
V_K(\bm{r}_i) &=& \frac{1}{4\pi\varepsilon_0} \int \frac{\rho_K(\mathbf{r}')}{|\mathbf{r}_i-\mathbf{r}'|} d^3\bm{r}' \nonumber \\
V_j(\bm{r}_i) &=& -\frac{1}{4\pi\varepsilon_0}\frac{e}{|\bm{r}_i-\bm{r}_j|}  \\
V_{L}(\bm{R}_K) &=& \frac{1}{4\pi\varepsilon_0} \frac{Z_L e}{|\bm{R}_K-\bm{R}_L|}. \nonumber
\end{eqnarray}
In Eq.~\eqref{eq:hamilelec}, as well as in all this work, the SI system of units was used. Here, the operator $V_K(\bm{r}_i)$ refers to the electrostatic potential produced by the nucleus $K$ at the position of the $i$-th electron, $V_j(\bm{r}_i)$ is the potential created by electron $j$ at the position $\bm{r}_i$, and $V_L(\bm{R}_K)$ is the electrostatic potential produced by nucleus $L$ at the position of nucleus $K$. $c$ is the speed of light in vacuum, $m_e$ is the electron rest mass, $e$ is the elementary charge, $\varepsilon_0$ is the permittivity of free space, $\rho_K(\mathbf{r}')$ is the charge density distribution of nucleus $K$ at an arbitrary position $\mathbf{r}'$, $Z_K$ and $Z_L$ are the atomic numbers of nucleus $K$ and $L$, respectively, and $\bm{r}_i$, $\bm{r}_j$, $\bm{R}_K$ and $\bm{R}_L$ are the position vectors of the electrons $i$ and $j$, and nuclei $K$ and $L$, respectively. Here and in what follows, the sums over $i$ and $j$ run over all the electrons in a molecule, whereas the sums over $K$ and $L$ run over its nuclei. $\hat{\bm{p}}_i$ is the linear momentum operator of electron $i$, while $\bm{\alpha}_i$ and $\beta_i$ are the $4\times 4$ Dirac matrices for this electron, and are expressed in the Dirac standard representation as
\begin{equation}\label{eq:ab}
    \bm{\alpha}=\begin{bmatrix}
    \emptyset_{2\times2} & \bm{\sigma}\\
    \bm{\sigma} & \emptyset_{2\times2}
    \end{bmatrix},
    \qquad
    \beta=\begin{bmatrix}
    \mathds{1}_{2\times2} & \emptyset_{2\times2}\\
    \emptyset_{2\times2} & -\mathds{1}_{2\times2}
    \end{bmatrix}.
\end{equation}
Here, $\bm{\sigma}=\sigma_x\,\bm{\hat{\i}}+\sigma_y\,\bm{\hat{\j}}+\sigma_z\,\bm{\hat{k}}$ is the Pauli vector, whereas $\mathds{1}_{2\times2}$ and $\emptyset_{2\times2}$ are the $2 \times 2$ identity and null matrices, respectively. The Pauli matrices $\sigma_x$, $\sigma_y$, and $\sigma_z$ are given by
\begin{equation}
    \sigma_x=\begin{bmatrix}
    0 & 1 \\
    1 & 0
    \end{bmatrix},
    \;\;\;\;
    \sigma_y=\begin{bmatrix}
    0      & -i \\
    i & 0
    \end{bmatrix},
    \;\;\;\;
    \sigma_z=\begin{bmatrix}
    1 & 0 \\
    0 & -1
    \end{bmatrix},
\end{equation}
\noindent with $i=\sqrt{-1}$ being the imaginary unit.

\subsection{eEDM enhancement factor \texorpdfstring{$W_\mathrm{d}$}{Wd}}\label{sec:wdtheory}

The effects on hydrogenic atoms arising from a $\mathcal{P,T}$-odd interaction between the permanent EDM of a single electron, parallel to its spin, and an electromagnetic field has been studied for the first time by Salpeter in 1958~\cite{Salpeter1958}. In his seminal work, he introduced a $\mathcal{P,T}$-odd perturbation term corresponding to a permanent eEDM into the one-electron Dirac equation in a Lorentz-covariant formulation. This term is analogous to the so-called ``Pauli moment'' interaction term (representing a QED interaction of the lowest order, for non-relativistic energies, between the electromagnetic field and the Pauli anomalous electric and magnetic dipole moments of the electron), but pre-multiplied by the pseudoscalar Dirac operator $\gamma^5$ (see pp.~47--51 of Ref.~\citenum{BetheSalpeter}).

In atoms or molecules with at least one electron whose spin is unpaired, the effects arising from these $\mathcal{P,T}$-violating interactions produce atomic or molecular (permanent) $\mathcal{P,T}$-violating electric dipole moments that may be significantly larger than that of a free electron. In the past it has been shown that the $\mathcal{P,T}$-violating EDM of these many-electron systems is mainly influenced by the electrostatic interactions between the eEDM of the unpaired electron and the internal electric fields~\cite{lindroth_1989_order}. Therefore, when calculating the molecular enhancement parameter $W_\mathrm{d}$ (see Eq.~\eqref{eq:HeffSR}) it is possible to ignore effects such as the interactions between the eEDM and the magnetic fields, and also the electron-electron Breit interactions.

When neglecting both, the interactions of the eEDMs with internal and external magnetic fields and the Breit interactions, the mean value of the Salpeter Hamiltonian can be equated to the expectation value of two different operators. One of these two effective Hamiltonians is the sum of only one-electron operators, while the other one also contains two-body operators. These two-electron contributions, however, have been shown to be considerably smaller than the one-electron ones in that particular effective Hamiltonian, and as such they can usually be safely neglected~\cite{sandars_1965_the,lindroth_1989_order}.

Therefore, by employing any of these two effective Hamiltonians (denoted henceforth as scheme 1 and scheme 2) to make theoretical predictions, one avoids having to treat the two-electron interactions of the Salpeter Hamiltonian, which are not negligible. The first of these effective Hamiltonians (i.e., within scheme 1) has the form
\begin{equation}\label{eq:HeEDMeff1}
    \hat{H}^\mathrm{eEDM}_{\textit{eff-}1}=-d_\mathrm{e}\sum_i(\beta_i-1)\bm{\Sigma}_i \cdot \bm{E}(\bm{r}_i),
\end{equation}
where the operator vectors $\bm{\Sigma}_i$ are related to the Pauli matrices by the expression
\begin{equation}
    \bm{\Sigma}=\begin{bmatrix}
    \bm{\sigma} & \emptyset_{2\times2} \\
    \emptyset_{2\times2} & \bm{\sigma}
    \end{bmatrix},
\end{equation}
and $\bm{E}(\bm{r}_i)$ is the total electrostatic electric field at the position of electron $i$, given by
\begin{equation}
    \bm{E}(\bm{r}_i)=-\bm{\nabla}_i \left[\sum_K V_K(\bm{r}_i) + V^{\text{ext}}(\bm{r}_i) + \sum_{j \neq i} V_j(\bm{r}_i) \right].
\end{equation}
Here, $V^{\text{ext}}(\bm{r}_i)$ is the electrostatic potential produced by an external electric field at the position of electron $i$, and $V_K(\bm{r}_i)$ and $V_j(\bm{r}_i)$ are the electrostatic potentials given in Eq.~\eqref{eq:Internal-potentials}.

In what follows, we neglect in Eq.~\eqref{eq:HeEDMeff1} the effects arising from the external electric fields and from the electric field produced by the electrons, as their contributions to the molecular enhancement factors have been shown to be small~\cite{sandars_1965_the,lindroth_1989_order,gaul_2019_systematic}. In this way, the effective Hamiltonian of scheme 1 can be expressed as
\begin{equation}\label{eq:HeEDMeff1-onlynuc}
    \hat{H}^\mathrm{eEDM}_{\textit{eff-}1} \approx -d_\mathrm{e}\sum_{K,i}(\beta_i-1)\bm{\Sigma}_i \cdot
    \left[-\bm{\nabla}_i V_K(\bm{r}_i)  \right] ,
\end{equation}
\noindent and if the nuclear charges are modelled using point-type density distributions, then this effective Hamiltonian reduces to the operator used throughout this work,
\begin{equation}\label{eq:HeEDMeff1-onlynuc-point}
    \hat{H}^\mathrm{eEDM}_{\textit{eff-}1} \approx -d_\mathrm{e}
    \sum_{K,i}
    \frac{Z_K e}{4\pi\varepsilon_0}
    (\beta_i-1)\bm{\Sigma}_i \cdot
    \frac{\bm{r}_i-\bm{R}_K}{|\bm{r}_i-\bm{R}_K|^3},
\end{equation}
\noindent where its dependence with the atomic number of the nuclei in the system is explicit.

The second effective eEDM Hamiltonian, within scheme 2, is given by~\cite{Martensson87,lindroth_1989_order}
\begin{equation}\label{eq:HeEDMeff2}
    \hat{H}^\mathrm{eEDM}_{\textit{eff-}2} = i \, d_\mathrm{e} \frac{2 c}{e\hbar}\sum_i\beta_i\gamma_i^5 \hat{p}_i^2,
\end{equation}
with $\gamma^5$ being the well-known $4 \times 4$ Dirac matrix defined as $\gamma^5=i \, \gamma^0\gamma^1\gamma^2\gamma^3$, where $\gamma^0=\beta$ and $\bm{\gamma}=\beta\bm{\alpha}=\gamma^1\,\bm{\hat{\i}}+\gamma^2\,\bm{\hat{\j}}+\gamma^3\,\bm{\hat{k}}$ (in terms of the Pauli matrices, $\gamma^0=\sigma_z \otimes \mathds{1}_{2\times2}$ and $\gamma^{1,2,3}= i \, \sigma_y \otimes \sigma_{x,y,z}$, so that $\gamma^5=\sigma_x \otimes \mathds{1}_{2\times2}$, with $\otimes$ implying a Kronecker product). Therefore, in the Dirac standard representation,
\begin{equation}
    \gamma^5=\begin{bmatrix}
    \emptyset_{2\times2}  & \mathds{1}_{2\times2} \\
    \mathds{1}_{2\times2} & \emptyset_{2\times2}
    \end{bmatrix}.
\end{equation}

Scheme 2 contains only one-electron operators within an approximation in which both the magnetic interactions and the electron-electron Breit interactions are neglected, reducing the computational complexity. However, a drawback of this effective Hamiltonian is in the fact that the non-relativistic limit of its mean value is not zero, while that of the Salpeter Hamiltonian vanishes~\cite{Schiff1963}. Furthermore, scheme 2 does not allow analysis of the separate nuclear contributions to the calculated $W_\mathrm{d}$ parameters.
Within scheme 1, on the other hand, the the non-relativistic limit of $W_\mathrm{d}$ is zero. While scheme 1 includes two-electron contributions (see Eq.~\eqref{eq:HeEDMeff1}), these are much smaller than the corresponding one-electron counterparts~\cite{lindroth_1989_order,gaul_2019_systematic} and  can usually be neglected. Therefore, scheme 1 allows us to study the effective contributions to $W_\mathrm{d}$ arising from each nucleus of the system, keeping the correct non-relativistic behavior, and using only one-electron operators.

The interactions of the eEDMs with the internal electric fields can be taken as a perturbation on the DC Hamiltonian $\hat{H}^{(0)}$ of Eq.~\eqref{eq:hamilelec}, and in such case the total (perturbed) electronic Hamiltonian is written as
\begin{equation}\label{eq:pertH}
\hat{H}^\mathrm{d}=\hat{H}^{(0)}+\lambda \, \hat{H}^\mathrm{eEDM},
\end{equation}
where $\lambda$ is the strength of the perturbation, and $\hat{H}^\mathrm{eEDM}$ can be either $\hat{H}_{\textit{eff-}1}^\mathrm{eEDM}$ or $\hat{H}_{\textit{eff-}2}^\mathrm{eEDM}$.

The corrections to the molecular electronic energy arising from the perturbed Hamiltonian can be obtained by using Rayleigh--Schrödinger perturbation theory. A series expansion of the ground-state energy solution of Eq.~\eqref{eq:pertH}, $E^\mathrm{d}_\Omega (\lambda)$, can be written around $\lambda=0$ as
\begin{equation}\label{eq:E-taylor}
E^\mathrm{d}_\Omega (\lambda)=E_\Omega^{(0)}+\lambda \, E_\Omega^{\mathrm{d}(1)}\Big|_{\lambda=0}+\frac{\lambda^2}{2} \, E_\Omega^{\mathrm{d}(2)}\Big|_{\lambda=0}+\mathcal{O}(\lambda^3),
\end{equation}
where the subindices $\Omega$ indicate that the ground-state solutions $| 0 \rangle$ of the unperturbed Hamiltonian $\hat{H}^{(0)}$ in a given fixed molecular frame and spin state fulfil the condition $\langle 0 | \hat{\Omega} | 0 \rangle = \Omega$. For YbCu, YbAg and YbAu, it can be seen that $|\Omega| = 1/2$~\cite{Tomza_2021}. Furthermore, the energy $E_\Omega^{(0)} = \langle 0 | \hat{H}^{(0)} | 0 \rangle$ is the eigenvalue of the unperturbed Hamiltonian $\hat{H}^{(0)}$ in the same molecular frame. When only the leading-order corrections are retained, and taking into account the relation between the effective spin-rotation Hamiltonian and the eEDM enhancement factor $W_\mathrm{d}$ given in Eq.~\eqref{eq:HeffSR}, it can be shown that this parameter is given as
\begin{equation}\label{eq:wd}
    W_\mathrm{d}
    = \frac{1}{\Omega \, d_\mathrm{e}} \left.\frac{dE^\mathrm{d}_{\Omega}}{d\lambda}\right|_{\lambda=0}
    = \frac{1}{\Omega \, d_\mathrm{e}} \left. \left( \frac{d}{d\lambda} \left\langle 0^\mathrm{d} \middle| \hat{H}^\mathrm{d} \middle| 0^\mathrm{d} \right\rangle \right) \right|_{\lambda=0},
\end{equation}
where $\left| 0^\mathrm{d} \right\rangle$ is the ground-state solution of the total (perturbed) Hamiltonian $\hat{H}^\mathrm{d}$.

\subsection{S-PS-ne enhancement factor \texorpdfstring{$W_\mathrm{s}$}{Ws}}\label{sec:wstheory}

A second source contributing to the $\mathcal{P,T}$-violating interactions involving the electric dipole moment of a polar paramagnetic molecule in a $^2\Sigma_\frac{1}{2}$ ground state are the $\mathcal{P,T}$-odd S-PS-ne neutral-current interactions. Assuming that in each nucleus $K$ the proton and neutron density distributions are equal to each other and also equal to the nuclear density distribution $\varrho_K(\bm{r})$, the effective Hamiltonian that accounts for the four-fermion semileptonic interactions in the electron–nucleon sector (in the limit of infinitely heavy nuclei) can be written in terms of the proton-electron and neutron-electron interaction constants $k_\mathrm{s}^\mathrm{p}$ and $k_\mathrm{s}^\mathrm{n}$, respectively, as~\cite{Gorshkov79,Kozlov85,Engel2013}
\begin{equation}\label{eq:HSPS}
    \hat{H}^\mathrm{S-PS-ne} = i \frac{G_F}{\sqrt{2}}
    \sum_{i,K} (Z_K k^\mathrm{p}_\mathrm{s} + N_K k^\mathrm{n}_\mathrm{s}) \, \beta_i\gamma^5_i\varrho_K(\bm{r}_{i}),
\end{equation}
\noindent where $N_K$ is the number of neutrons in nucleus $K$, $G_F$ is the Fermi coupling constant (whose most recent value is $G_F/(\hbar \, c)^3=1.1663787 \times 10^{-5}$~GeV$^{-2}$, or equivalently $G_F\simeq 2.222516 \times10^{-14} \, E_h \, a_0^3$~\cite{codata2018}), and $\varrho_K(\bm{r}_{i})= \rho_K(\bm{r}_{i})/(Z_K e)$ is the nuclear density distribution of nucleus $K$ at the position of the $i$-th electron, satisfying $\int \varrho_K (\mathbf{r}) \, d^3\mathbf{r} = 1$.
Besides, $E_h$ and $a_0$ are the Hartree energy and Bohr radius, respectively.
%
By defining a factor $k_{\mathrm{s},K}= k^\mathrm{p}_\mathrm{s} + (N_K/Z_K) \, k^\mathrm{n}_\mathrm{s}$, we can rewrite Eq.~\eqref{eq:HSPS} as
\begin{eqnarray}\label{eq:HSPS-2}
    \hat{H}^\mathrm{S-PS-ne} &=&
    \sum_K \hat{H}_K^\mathrm{S-PS-ne} \nonumber \\
    &=&
    i \frac{G_F}{\sqrt{2}}
    \sum_{i,K} Z_K k_{\mathrm{s},K} \, \beta_i\gamma^5_i\varrho_K(\bm{r}_{i}).
\end{eqnarray}

Making a treatment analogous to the one applied to the eEDM Hamiltonian, it is easy to see that the S-PS-ne Hamiltonian can also be taken as a perturbation (with field strength $\lambda$) to the 4C DC Hamiltonian, so that
\begin{equation}\label{eq:pertHWs}
\hat{H}^\mathrm{s}=\hat{H}^{(0)}+\lambda\,\hat{H}^\mathrm{S-PS-ne}.
\end{equation}
By expanding the solution energy around $\lambda=0$, the enhancement factor $W_\mathrm{s}$ can be obtained as
\begin{equation}\label{eq:ws}
W_\mathrm{s}=\frac{1}{\Omega} \sum_K \frac{1}{k_{\mathrm{s},K}}\left.\frac{dE^{\mathrm{s},K}_\Omega}{d\lambda}\right|_{\lambda=0}.
\end{equation}

\section{Computational details}\label{sec:comp-det}

The calculations were performed using the DIRAC-19.0 program package~\cite{DIRAC19,dirac-paper}, in the framework of the 4C DC Hamiltonian. The multi-reference Fock-space coupled-cluster (FSCC) method with single and double excitations was used to treat electron correlation effects~\cite{visscher_2001_formulation}. A multi-reference method is required due to the challenging character of the ytterbium-containing molecules~\cite{DenHaoEli20,ZhaZhaChe22}. This method was employed previously to study the enhancement factors of the $\mathcal{P,T}$-violating interactions in YbOH~\cite{denis_2019_enhancement} and YbCH$_3$~\cite{chamorro_2022_molecular}. Additionally, we also used the FSCC implementation within the EXP-T program~\cite{oleynichenko_2020_EXP,oleynichenko_2020_towards} to investigate the effects of including triple excitations.

The uncontracted Dyall's valence-only basis sets of double-$\zeta$ (v2z), triple-$\zeta$ (v3z) and quadruple-$\zeta$ (v4z) quality were employed~\cite{dyall_2004_relativistic,dyall_2007_relativistic,dyall_2009_revised,gomes_2010_relativistic,dyall_2022_diffuse}. Furthermore, the core-valence basis sets (cv$X$z, $X=2,3,4$) were also used in order to examine the effect of correlating the core electrons~\cite{dyall_2011}. These particular basis sets add tight functions with large exponents. The augmented basis sets (s-aug-v$X$z) were employed to investigate how accurately the outer regions of the systems were described. These basis sets add a single diffuse function to each symmetry block.

\subsection{Finite-field method}

The molecular enhancement factors $W_\mathrm{d}$ and $W_\mathrm{s}$ were calculated by employing the finite-field method. In particular, by combining Eqs.~\eqref{eq:E-taylor}, \eqref{eq:wd} and \eqref{eq:ws} it can be seen that they can be obtained by applying the two-point finite-field method, where
\begin{equation}\label{eq:Wd-ff}
W_\mathrm{d} \approx \frac{1}{\Omega \, d_\mathrm{e}} \left[ \frac{E^\mathrm{d}_\Omega (\lambda) - E^\mathrm{d}_\Omega (-\lambda)}{2\lambda} \right], 
\end{equation}
and
\begin{equation}\label{eq:Ws-ff}
W_\mathrm{s} \approx \frac{1}{\Omega} \sum_K \frac{1}{k_{\mathrm{s},K}} \left[ \frac{E^{\mathrm{s},K}_\Omega (\lambda) - E^{\mathrm{s},K}_\Omega (-\lambda)}{2\lambda} \right].
\end{equation}

A field strength $\lambda=10^{-6}$ was set when studying both parameters $W_\mathrm{d}$ and $W_\mathrm{s}$ in YbCu and YbAg, whereas $\lambda=10^{-7}$ was used for the calculations involving YbAu.

Atomic units (i.e., $e=1$, $a_0=1$, $\hbar=1$, and $4\pi\varepsilon_0=1$) were used in all the calculations. For $W_\mathrm{d}$, the values obtained following Eq.~\eqref{eq:Wd-ff} were converted to the units used throughout this work by means of a conversion factor equal to the atomic unit (a.u.) of electric field $E_h/(e\,a_0)=1.243380059\times10^{24}\;\frac{h\,\text{Hz}}{e\,\text{cm}}$. To calculate $W_\mathrm{s}$ following Eq.~\eqref{eq:Ws-ff}, the energies $E^{\mathrm{s},K}_\Omega (\pm\lambda)/k_{\mathrm{s},K}$ were obtained in a.u.~as mean values of the Hamiltonian
\begin{equation}
\hat{H}^{(0)} \pm \frac{\lambda}{k_{\mathrm{s},K}} \frac{\sqrt{2}}{G_F\,Z_K}\hat{H}^\mathrm{S-PS-ne}_K,
\end{equation}
and then the factor $G_F\,Z_K/(\sqrt{2}\,a_0^3)=Z_K\,0.103403426\,h\,\unit{\kilo\hertz}$ was used to convert the values of $W_\mathrm{s}(K)$ to $h\,\unit{\kilo\hertz}$. All values of the fundamental constants were taken from Ref.~\citenum{codata2018}.

\subsection{Geometry optimization}\label{sec:bondlength}

The enhancement factors were computed for the equilibrium geometries of the systems. The bond lengths were determined using the exact two component (X2C) Hamiltonian~\cite{ilia_2007_an}, the single-reference CCSD method, and s-aug-v4z basis sets. By employing the X2C Hamiltonian, the Dirac equation is transformed decoupling the large and small components of the Dirac spinors. This method only takes positive energy solutions into account. The active space energy cut-offs for the virtual (unoccupied) and occupied orbitals were set to $\pm20\,E_h$, $\pm10\,E_h$ and $\pm10\,E_h$ for YbCu, YbAg and YbAu, respectively. 
A smaller active space was used for the heavier molecules since computing their bond distances is computationally more intensive.
The results of these geometry optimizations can be found in Table \ref{tab:distances} along with the values obtained previously in Ref.~\citenum{Tomza_2021}.
\begin{table}[bt]
\centering
\caption{Equilibrium bond distances, $R_{e}$, of YbCu, YbAg and YbAu ({\AA}) computed at the X2C/CCSD/s-aug-v4z level of theory. Including comparison to previous results obtained using energy-consistent pseudopotentials (ECP).}
\label{tab:distances}
\renewcommand*{\arraystretch}{1.3}
\begin{tabular*}{\linewidth}{@{\extracolsep{\fill}} lll @{\hskip 10pt} lll}
 \toprule
  & &&\multicolumn{3}{c}{$R_{e}$ [\r{A}]}\\\cmidrule(lr){4-6}
 Source & Method & Relativity & YbCu & YbAg & YbAu \\[0.5ex]
 \midrule
 This work & CCSD  & X2C & 2.7543 & 2.8589 & 2.6524\\
 Ref.~\citenum{Tomza_2021} & CCSD(T)  & ECP & 2.910 & 3.063 & 2.939\\
 \bottomrule
\end{tabular*}
\end{table}

The general trend observed for both sets of results is that the bond length increases from YbCu to YbAg and then decreases from YbAg to YbAu, likely due to the relativistic contraction of the 6$s$ orbital of gold. A non-relativistic treatment would show longer bonds for heavier systems~\cite{pyykko_1988_relativistic}.

The discrepancy between the present and the previous values is likely due to the use of pseudopotentials (especially a large-core one with a core polarization potential for Yb) in the latter. ECPs are limited to scalar-relativistic effects, while the X2C procedure employed here also accounts for the spin-orbit coupling. The fact that the largest discrepancy is found for the heavier system supports this assumption.

\section{Results and discussion}\label{sec:results}
 
To obtain accurate enhancement factors accompanied by well-defined uncertainties, multiple computational aspects will be addressed. For all the calculations discussed in this section, the 4C DC Hamiltonian was employed. First, the baseline results will be presented in Section~\ref{sec:baseline}. Then, the effect of selecting two different nuclear charge density distribution models is examined for both factors in Section~\ref{sec:nucsize}. Furthermore, the two schemes employed for computing $W_\mathrm{d}$ will be compared in Section~\ref{sec:schemes}. Next, the influence of the basis set will be determined in Section~\ref{sec:basisset}. Thereafter, different computational approaches will be compared in Section~\ref{sec:methods}. Finally, the effect of the geometry of the system on $W_\mathrm{d}$ and $W_\mathrm{s}$ will be discussed in Section~\ref{sec:geometry}.

This extensive investigation allows us to set uncertainties on the recommended values. The justification for the final results and their uncertainties will be given in Section~\ref{sec:final}.

\subsection{Baseline calculations}\label{sec:baseline}

All the reference values for $W_\mathrm{d}$ and $W_\mathrm{s}$ were computed on the v3z/FSCCSD level. For YbCu and YbAg, a virtual space cut-off of $500\,E_h$ was employed, and 2 and 4 electrons were frozen, respectively. For YbAu, the virtual space cut-off was set to $40\,E_h$, and 56 electrons were frozen. The selection of these correlation parameters is justified in the Supplementary Material.

The calculated enhancement factors for the three systems are given in Table \ref{tab:baselineresults}. We can observe that both parameters are very similar for YbCu and YbAg, while much lower absolute values are obtained for YbAu. This finding will be elucidated in the following sections. Furthermore, multiple corrections to these baseline values will be determined. The final obtained values will be given in Section~\ref{sec:final}.

\begin{center}
\begin{table}[bt]
\centering
\caption{Reference values of $W_\mathrm{d}$ and $W_\mathrm{s}$ for YbCu, YbAg and YbAu. These results were obtained on the FSCCSD/v3z level of theory, while freezing 2, 4 and 56 electrons, respectively.}
\label{tab:baselineresults}
\renewcommand*{\arraystretch}{1.3}
\begin{tabular}{l@{\hskip 25pt} l @{\hskip 20pt} l } 
 \toprule
 Molecule & $ W_\mathrm{d} \, \bigl[10^{24}\frac{h\,\text{Hz}}{e\,\text{cm}}\bigr]$ & $ W_\mathrm{s} \, [h\,\text{kHz}]$\\
 \midrule
 YbCu & 13.122 & --47.647\\
 YbAg & 11.869 & --44.361\\
 YbAu & \phantom{0}1.326 & \phantom{--0}6.979\\
 \bottomrule
\end{tabular}
\end{table}
\end{center}

\subsection{Nuclear size effects}\label{sec:nucsize}

The effect of the nuclear model on both molecular enhancement factors was investigated. For all three molecules, $W_\mathrm{d}$ and $W_\mathrm{s}$ were calculated by employing both a point-like and a spherically symmetric Gaussian-type function to model the nuclear charge densities of each nucleus $K$, $\rho_K(\bm{r})$, as well as the normalized densities $\varrho_K(\bm{r}) = \rho_K(\bm{r})/(Z_K e)$ appearing in $\hat{H}^\mathrm{S-PS-ne}$ (see Eq.~\eqref{eq:HSPS}). The two type of charge densities (point nucleus, PN, and Gaussian-type nucleus, GN) employed in this work can be expressed as
\begin{eqnarray}
\rho^{PN}_K(\bm{r}) &=& Z_K e \, \delta(\bm{r}-\bm{R}_K)\nonumber \\
\rho^{GN}_K(\bm{r}) &=& Z_K e \left(\frac{\zeta_K}{\pi}\right)^{3/2} e^{-\zeta_K \; |\bm{r} - \mathbf{R}_K|^2},
\end{eqnarray}
where $\delta(\bm{r})$ is the Dirac delta distribution and $\zeta_K = 3/(2\langle R_{\text{nuc},K}^2\rangle)$, with $\sqrt{\langle R_{\text{nuc},K}^2 \rangle}$ being the root-mean-square radius of the nucleus $K$, which can be obtained using the empirical relation $\sqrt{\langle R_{\text{nuc},K}^2 \rangle}=(0.836 \, A_K^{1/3}+0.570)\,\text{fm}$~\cite{Andrae2000}, where $A_K$ is the mass number of the isotope of interest.

For this analysis, all calculations were performed on the FSCCSD/v2z level of theory, correlating all electrons and using a virtual space cut-off of $3000\,E_h$ for all three systems.

\begin{figure}[bt]
    \centering
    \includegraphics[width=.45\textwidth]{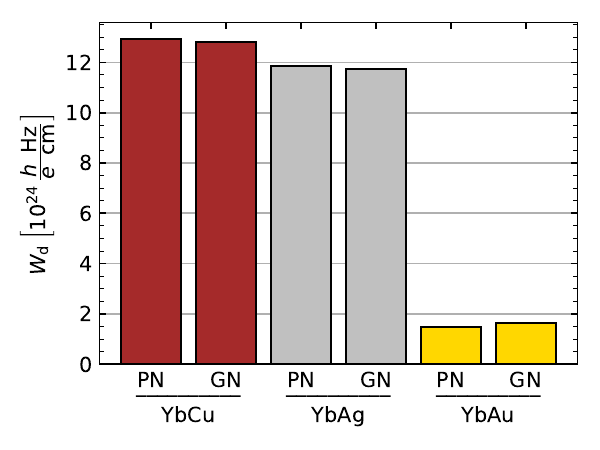}
    \caption{Graphical representation of the behavior of $W_\mathrm{d}$ for YbCu, YbAg and YbAu using Gaussian-type and point-type nuclear models (GN and PN, respectively). These results were obtained on the FSCCSD/v2z level, correlating all the electrons, and with a symmetric virtual space cut-off.}
    \label{fig:nucsize_wd}
\end{figure}

Figure~\ref{fig:nucsize_wd} shows the effect of the nuclear model on $W_{\mathrm{d}}$. For YbCu and YbAg, the results using GN are $\sim1.1\%$ smaller than those obtained employing PN, while for YbAu the GN result is $\sim9.2\%$ larger than the one using a PN.

\begin{figure}[bt]
    \centering
    \includegraphics[width=.45\textwidth]{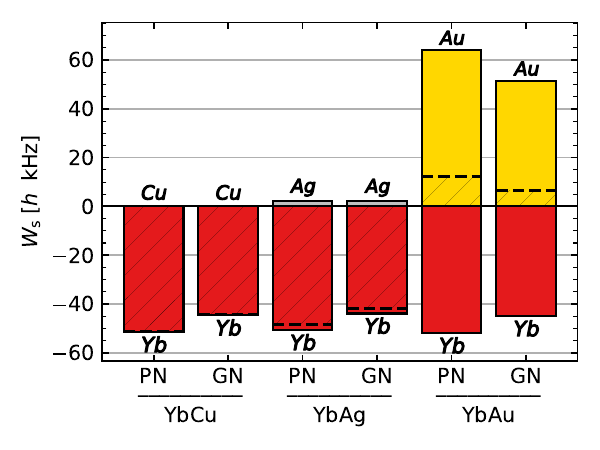}
    \caption{Contributions to $W_\mathrm{s}$ for YbCu, YbAg and YbAu, arising from each individual nucleus. The results were obtained on the FSCCSD/v2z level of theory with all electrons correlated and a symmetric virtual space cut-off and using Gaussian-type and point-like nuclear models (GN and PN, respectively). The hatched blocks correspond to the sum $W_\mathrm{{s},\text{Yb}}+W_{\mathrm{s},X}$ ($X=$ Cu, Ag, Au).}
    \label{fig:nucsize_ws}
\end{figure}

Figure~\ref{fig:nucsize_ws} shows the nuclear size effects on $W_\mathrm{s}$. The contributions from each of the nuclei are shown, as well as the total values of $W_\mathrm{s}$, represented by hatched blocks. The contributions associated with the nuclei of the coinage metals become increasingly important as their atomic number increases. Moreover, as expected, these contributions are of opposite sign to those of ytterbium, because in the region between the two nuclei the gradients of the nuclear densities have opposite directions. For YbAu, the two contributions almost cancel each other out, leading to a very small total value of $W_\mathrm{s}$.

In all three cases, the contribution of the ytterbium nucleus is reduced by $\sim13\%$ when going from PN to GN, while for copper, silver, and gold the reduction is $\sim$~$0.16\%$, $\sim3.5\%$, and $\sim20\%$, respectively. Thus, the effect of using a finite nucleus model becomes more significant for the heavier elements, as can be expected. The total absolute $W_\mathrm{s}$ is also lower for the calculations performed using the GN model. 

\subsection{$W_\mathrm{d}$: Comparison of schemes 1 and 2}\label{sec:schemes}

It is well known that when only the electric field produced by the nuclei is taken into account in scheme 1, the two schemes described in Section~\ref{sec:wdtheory} should yield similar results for $W_\mathrm{d}$~\cite{lindroth_1989_order}. While the use of scheme 2 is computationally less demanding, since it requires just a single calculation per system instead of the two that are required for each diatomic molecule, scheme 1 allows us to examine the effective contributions arising from each nucleus. In order to study those individual contributions, in this work we present (to the best of our knowledge) the first four-component results of $W_\mathrm{d}$ using the approximate effective Hamiltonian of Eq.~\eqref{eq:HeEDMeff1-onlynuc-point}, corresponding to the use of scheme 1.

The enhancement factors $W_\mathrm{d}$ were computed using both schemes for the three systems considered in this work, on the FSCCSD/v2z level of theory, with symmetric cut-offs of $500\,E_h$, $500\,E_h$, and $40\,E_h$, freezing 2, 4, and 56 electrons of YbCu, YbAg and YbAu, respectively. The calculations were only performed using a PN model (and not a GN nucleus), since some integrals required for scheme 1 for finite nuclear models, despite having been studied using two-component methods~\cite{gaul_2019_systematic}, are currently not implemented in DIRAC.

\begin{figure}[bt]
    \centering
    \includegraphics[width=.45\textwidth]{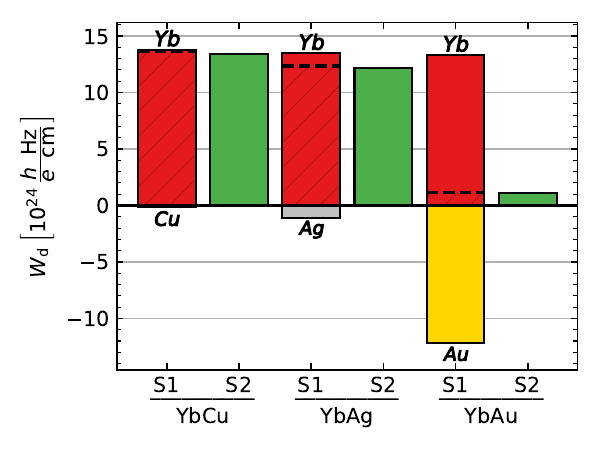}
    \caption{Graphical representation of the behavior of $W_\mathrm{d}$ of YbCu, YbAg and YbAu computed using scheme 1 (S1) and scheme 2 (S2). For S1, the contributions from both nuclei are shown as well as their sums (hatched blocks). These results were obtained on the FSCCSD/v3z level with a symmetric virtual space cut-off.}    
    \label{fig:schemes}
\end{figure}

Figure~\ref{fig:schemes} shows the difference in $W_\mathrm{d}$ when computed with the two schemes for the three systems. The total YbCu $W_\mathrm{d}$ calculated using scheme 2 is $\sim1.5\%$ smaller than the scheme 1 result. For YbAg, this difference is $\sim1.4\%$. For YbAu, the reduction is $\sim4.7\%$.

Using scheme 1, it can be seen that the contributions to $W_\mathrm{d}$ from the Yb nucleus remain almost constant for all three systems. Moreover, the contributions from the coinage metals all have opposite signs to those coming from Yb, as is the case for the $W_\mathrm{s}$ factors. This is expected, since the electric fields due to the two nuclei in the internuclear region have opposite directions, and according to Eqs.~\eqref{eq:HeEDMeff1-onlynuc} and \eqref{eq:HeEDMeff1-onlynuc-point}, this generates opposite contributions to this enhancement parameter. The decreasing total $W_\mathrm{d}$ factor from YbCu to YbAu is due to the increasing contribution from the second nucleus (opposite to that of the first).

\subsection{Basis set effects}\label{sec:basisset}

To observe how the size of the basis set influences the enhancement factors, $W_\mathrm{d}$ and $W_\mathrm{s}$ were computed with double-$\zeta$, triple-$\zeta$ and quadruple-$\zeta$ quality basis sets. These calculations were done using the FSCCSD method, and the (occupied and virtual) active space cut-offs were set to $\pm20\,E_h$, $\pm10\,E_h$ and $\pm10\,E_h$, freezing 38, 64 and 82 electrons of YbCu, YbAg and YbAu, respectively.

\begin{figure}[bt]
    \centering
    \includegraphics[width=.45\textwidth]{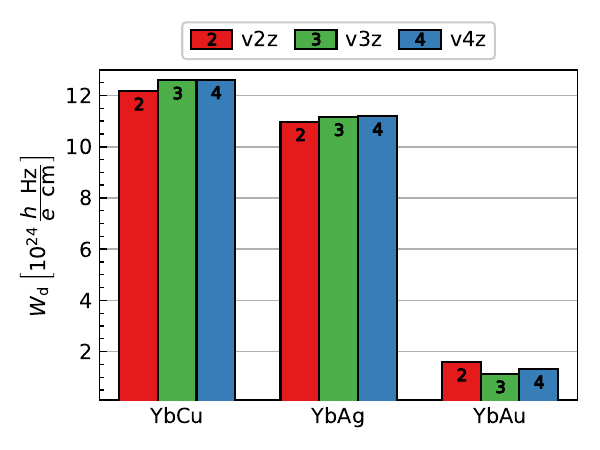}
    \caption{The $W_\mathrm{d}$ enhancement factor of YbCu, YbAg and YbAu, for v2z, v3z and v4z basis sets. Computations were performed at the FSCCSD level.}
    \label{fig:basis_set_quality}
\end{figure}

\begin{figure}[bt]
    \centering
    \includegraphics[width=.45\textwidth]{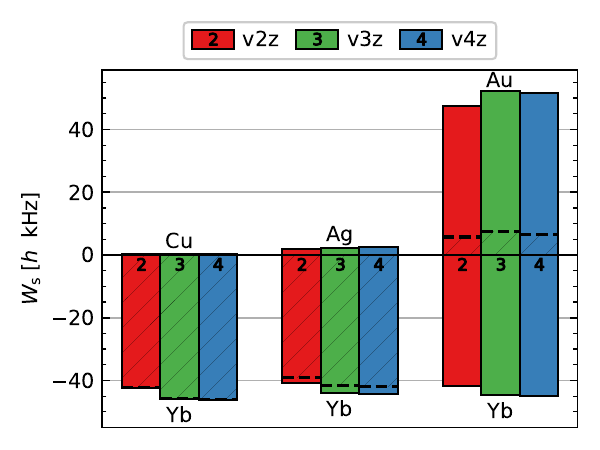}
    \caption{The $W_\mathrm{s}$ enhancement factor of YbCu, YbAg and YbAu, for v2z, v3z and v4z basis sets, individual contributions from each nucleus. Computations were performed at the FSCCSD level. The hatched blocks correspond to the sum of the contributions from both elements.}
    \label{fig:basis_set_quality_Ws}
\end{figure}

The plot in Figure~\ref{fig:basis_set_quality} shows the effect of increasing the basis set cardinality on $W_\mathrm{d}$. Converging behavior can be observed for parameter values with increasing basis set quality for YbCu and YbAg. While no apparent convergence can be observed for the total $W_\mathrm{s}$ values of YbAu, this convergence can be seen by looking at the individual contributions in Figure \ref{fig:basis_set_quality_Ws}. A similar trend is expected for $W_\mathrm{d}$.

Apart from adding contributing functions to all orbitals, it is also possible to add only tight or diffuse functions. Tight functions should increase the accuracy of the description of the core region of the system, while diffuse functions should improve the accuracy of the description of the valence region of the system. The cv$X$z basis sets contain higher angular momentum tight functions, and the s-aug-v$X$z basis sets augment a diffuse function to the v$X$z basis set. These computations were done using the same computational settings as the ones above.

The effects of increasing the accuracy on the description of the core and valence regions of the systems is small for both enhancement factors. Adding tight functions has a negligible effect on all three systems for $W_\mathrm{d}$ (with the largest change of $0.005 \times 10^{24}\frac{h\,\text{Hz}}{e\,\text{cm}}$) and $W_\mathrm{s}$ (with the largest change of $0.024~h\,\text{kHz}$). On the other hand, adding diffuse functions reduces the parameter values slightly (with the largest changes of $0.033 \times 10^{24}\frac{h\,\text{Hz}}{e\,\text{cm}}$ and $0.143~h\,\text{kHz}$ for $W_\mathrm{d}$ and $W_\mathrm{s}$, respectively). Furthermore, the enhancement factors of the YbAg system are more affected than those of YbCu. For YbAu, the addition of diffuse functions increases the calculated $W_\mathrm{d}$ and decreases the $W_\mathrm{s}$. Still, the relative effects  remain within $3\%$ of the total value, as can be seen in the Supplementary Material.

For the final calculations, the v3z basis set was used. For all the calculated enhancement factors, the differences between the use of v4z and v3z basis sets is small compared to the electron correlation effects. The incompleteness of the basis set will be taken into account in the uncertainty.

\subsection{Computational methods}\label{sec:methods}

The method used so far throughout this work for calculations of the enhancement factors is 4C Dirac--Coulomb FSCC.

\begin{figure}[bt]
    \centering
    \includegraphics[width=.45\textwidth]{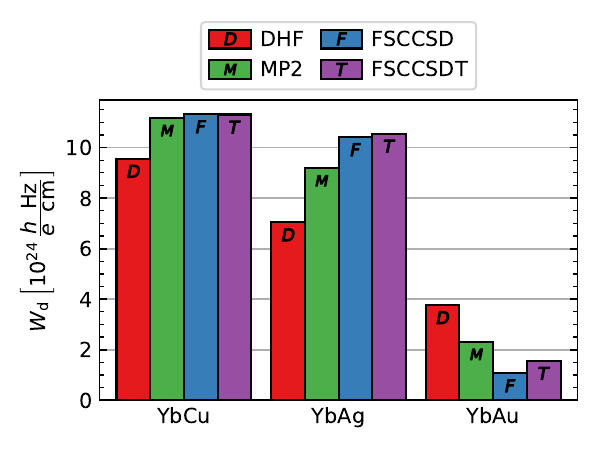}
    \caption{The $W_\mathrm{d}$ enhancement factor of YbCu, YbAg and YbAu computed at DHF, MP2, and FSCC levels of approach. These computations were performed using the v2z basis set, and the active space cut-offs were set to $\pm2\,E_h$, freezing 66, 84 and 116 electrons of YbCu, YbAg and YbAu, respectively.}
    \label{fig:wdmethods}
\end{figure}

\begin{figure}[bt]
    \centering
    \includegraphics[width=.45\textwidth]{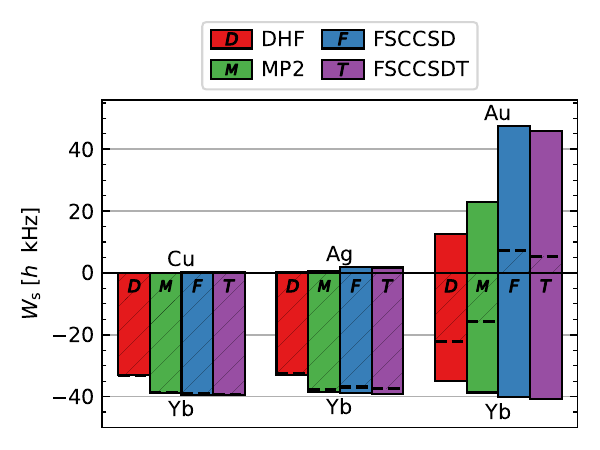}
    \caption{The $W_\mathrm{s}$ enhancement factor of YbCu, YbAg and YbAu computed at the level of DHF, MP2, and FSCC approaches. Results from each nucleus are given separately. The combined values are indicated by the hatched blocks. These computations were performed using the same basis set and active space cut-offs as in Fig.~\ref{fig:wdmethods}.}
    \label{fig:wsmethods}
\end{figure}

In Table \ref{tab:methods} and Figures \ref{fig:wdmethods} and \ref{fig:wsmethods}, different approaches are compared: DHF, Møller–Plesset up to second order (MP2), FSCCSD and FSCCSDT. These calculations were performed employing the dyall.v2z basis set, and the active space cut-offs for the post-DHF computations were set to $\pm2\,E_h$, freezing 66, 84 and 116 electrons of YbCu, YbAg and YbAu, respectively. This small active space was chosen because inclusion of triple excitations (using the EXP-T program) is computationally expensive. The DHF results differ the most from the FSCC values, while the MP2 values are close to the CC results for YbCu and YbAg. On the other hand, going from MP2 to FSCC reduced the value of $W_\mathrm{d}$ for YbAu by a factor of two and reverses the sign of $W_\mathrm{s}$.

\begin{center}
\begin{table}[bt]
\caption{The enhancement factors $W_\mathrm{d}$ and $W_\mathrm{s}$ of YbCu, YbAg and YbAu computed at the DHF, MP2, and two different FSCC levels of approach. Computations were performed using v2z basis set.}
\label{tab:methods}
\renewcommand*{\arraystretch}{1.3}
\centering
\begin{tabular}{l @{\hskip 10pt} lll @{\hskip 10pt} lll} 
 \toprule
  & \multicolumn{3}{c}{$W_\mathrm{d}\,\bigl[10^{24}\frac{h\,\text{Hz}}{e\,\text{cm}}\bigr]$} & \multicolumn{3}{c}{\;\;$W_\mathrm{s}\,[h\,\text{kHz}]\;\;$} \\\cmidrule(lr{0.2in}){2-4}\cmidrule(lr){5-7}

 \rule{0pt}{2.5ex}Method & YbCu & YbAg & YbAu & \;YbCu & YbAg & YbAu \\[0.5ex]
 \midrule
 DHF & \phantom{0}9.652 & \phantom{0}7.044 & \phantom{0}3.770 & --33.081 & --32.547 & --22.275\\
 
 MP2 & 11.174 & \phantom{0}9.188 & \phantom{0}2.314 & --38.670 & --37.696 & --15.822\\
 
 FSCCSD & 11.323 & 10.415 & \phantom{0}1.072 & --39.162 & --36.897 & \phantom{--0}7.385\\
 
 FSCCSDT & 11.310 & 10.555 & \phantom{0}1.567 & --39.210 & --37.406 & \phantom{--0}5.268\\
 \bottomrule
\end{tabular}
\end{table}
\end{center}

Including the triple excitations in the FSCC calculations has only a minor effect on the enhancement factors of YbCu and YbAg, but increases the $W_\mathrm{d}$ and decreases the $W_\mathrm{s}$ of YbAu significantly, in line with enhanced sensitivity of this systems to the other computational parameters. The  differences between the FSCCSD and FSCCSDT methods will be used to estimate the uncertainty due to neglect of the higher excitations in Section~\ref{sec:final}.

\subsection{Influence of molecular geometry}\label{sec:geometry}

\subsubsection{Uncertainty associated to the bond length}

The equilibrium bond lengths used in the calculations of the enhancement factors are given in Section~\ref{sec:bondlength}. The accuracy of these theoretically predicted values compared to experiments is unknown, since no experimental bond lengths are available for the systems considered in this work. Since the bond length has a significant impact on the studied enhancement factors, the uncertainty associated to the bond length should be taken into account. The difference between the calculated equilibrium bond lengths obtained within the FSCCSD approach and the experimental values is usually on the order of \SI{0.01}{\angstrom}~\cite{LadjimiPRA24}. 

All enhancement factors were computed at the calculated equilibrium bond distances, as well as at \SI{0.01}{\angstrom} larger and smaller internuclear distances. These computations were performed on the FSCCSD/v2z level of theory, and the (occupied and virtual) active space cut-offs were set to $\pm100\,E_h$, $\pm100\,E_h$ and $\pm95\,E_h$, freezing 12, 20 and 28 electrons of YbCu, YbAg and YbAu, respectively. The results are given in Table~\ref{tab:runcertainty}.

\begin{center}
\begin{table}[bt]
\caption{The enhancement factors $W_\mathrm{d}$ and $W_\mathrm{s}$ of YbCu, YbAg and YbAu, computed at three different displacements ($\delta_R$) with respect to the equilibrium bond length. Computations were performed on the FSCCSD/v2z level of approach.}
\label{tab:runcertainty}
\renewcommand*{\arraystretch}{1.3}
\centering
\begin{tabular}{l @{\hskip 20pt} lll @{\hskip 15pt} lll} 
 \toprule
  & \multicolumn{3}{c}{$W_\mathrm{d}\,\bigl[10^{24}\frac{h\,\text{Hz}}{e\,\text{cm}}\bigr]$} & \multicolumn{3}{c}{\;\;$W_\mathrm{s}\,[h\,\text{kHz}]\;\;$} \\\cmidrule(lr{0.3in}){2-4}\cmidrule(lr){5-7}
 \rule{0pt}{2.5ex}$\delta_R$ [\unit{\angstrom}] & YbCu & YbAg & YbAu & \;YbCu & YbAg & YbAu \\[0.5ex]
 \midrule
 --0.01 & 12.536 & 11.494 & 1.541 & --43.476 & --40.881 & 6.581\\
 
 \phantom{--}0.00 & 12.512 & 11.477 & 1.619 & --43.390 & --40.813 & 6.219\\

 \phantom{--}0.01 & 12.487 & 11.460 & 1.695 & --43.303 & --40.742 & 5.866\\
 
 \bottomrule
\end{tabular}
\end{table}
\end{center}

For both enhancement factors of YbCu and YbAg, the result decreases by only $0.1-0.2\%$ for the larger internuclear distance and increases only by $0.1-0.2\%$ for the smaller internuclear distance. For YbAu, the deviation is significantly larger, at $5.0-7.0\%$.

\subsubsection{Vibrational corrections}\label{sec:vibcorr}
The anharmonicity of the potential energy curve for the electronic ground state implies that the effective equilibrium bond distance is shifted slightly compared to the minimum of the potential energy curve. This slight difference in bond length results in small changes in the enhancement factors.

\begin{figure}[bt]
    \centering
    \includegraphics[width=.45\textwidth]{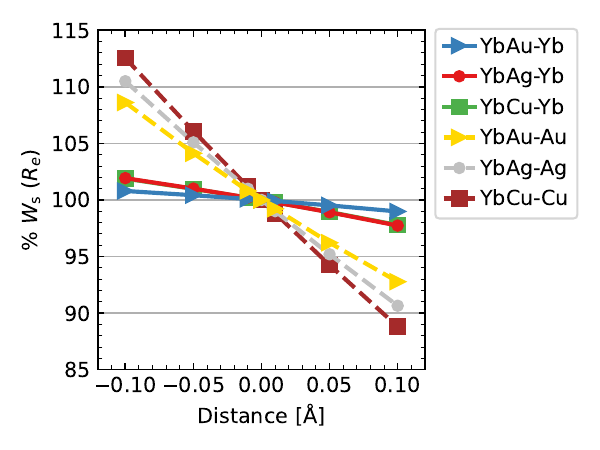}
    \caption{The effect of the internuclear distance on the enhancement factor $W_\mathrm{s}$ of YbCu, YbAg and YbAu. These results were obtained on the FSCCSD/v2z level.}
    \label{fig:geometry}
\end{figure}

To compute the vibrational correction, calculations of the enhancement factors and the potential energies for different bond lengths were performed. The results for $W_\mathrm{s}$, split up into the contribution arising from each nucleus, can be found in Figure~\ref{fig:geometry}. These values are given as a function of the difference between the internuclear distances and the equilibrium bond length of the corresponding molecule (in Angstroms). The resulting $W_\mathrm{s}$ values are given as a percentage compared to the results found at the equilibrium bond length.
It can be seen that the coinage metal contribution is significantly more sensitive to the bond length effect compared to the Yb contribution. This is also in accordance to the sensitivity of the respective atoms to the description of the electronic structure, as can be seen in Figure~\ref{fig:wsmethods}.

The correction values were computed using the VIBCAL module in DIRAC-19.0, using a fourth-order polynomial for the energy fitting. The vibrational corrections are listed in Table~\ref{tab:geometry} for both enhancement factors.

\begin{table}[bt]
\centering
\caption{The vibrational correction for all enhancement factors. These results were obtained on the FSCCSD/v2z level of theory.}
\label{tab:geometry}
\renewcommand*{\arraystretch}{1.3}
\begin{tabular}{l@{\hskip 25pt} l @{\hskip 20pt} l }
 \toprule
 Molecule & $\Delta W_\mathrm{d}\,\bigl[10^{24}\frac{h\,\text{Hz}}{e\,\text{cm}}\bigr]$ & $\Delta W_\mathrm{s}\,[h\,\text{kHz}]$\\
 \midrule
 YbCu & --0.0322 & --0.1154\\
 YbAg & --0.0235 & --0.0857\\
 YbAu & \phantom{--}0.0130 & \phantom{--}0.0479\\
 \bottomrule
\end{tabular}
\end{table}

The vibrational corrections alter the original values (obtained at $R_e$) at most by $\sim1\%$. These differences will also be taken into account when computing the final values in Section~\ref{sec:final}.

\subsection{Final recommended values and uncertainties}\label{sec:final}

The baseline results given in Section~\ref{sec:baseline} can now be corrected to obtain the final recommended values of $W_\mathrm{d}$ and $W_\mathrm{s}$. The contributions resulting from using a larger basis set (dyall.v4z), correlating all electrons, increasing the virtual space cut-offs, including triple excitations, and taking into account vibrational corrections were added to the reference baseline values. Table~\ref{tab:finalresults} presents these different contributions and also the final recommended values.

In particular, the basis set corrections were calculated as differences between FSCCSD calculations with active space cut-offs of $\pm20\,E_h$ for YbCu and $\pm10\,E_h$ for YbAg and YbAu, employing the dyall.v4z and dyall.v3z basis sets (see Table VI of the Supplementary Material). 

The effects arising from correlating all electrons were taken as the differences between freezing 2, 4 and 56 electrons for YbCu, YbAg and YbAu respectively and correlating all the electrons and simultaneously increasing the virtual cutoff to $6000\,E_h$ at the FSCCSD/dyall.v2z level of theory. A detailed analysis of the effects of the active correlation space size is available in the Supplementary Material.

The effects due to the inclusion of higher excitations were taken as the differences between FSCCSDT and FSCCSD calculations, employing v2z basis sets and cut-offs of $\pm2\,E_h$. Finally, the vibrational effects were extracted from Table~\ref{tab:geometry}.

\begin{table*}[bt]
\centering
\caption{Final recommended values of $W_\mathrm{d}$ and $W_\mathrm{s}$ for YbCu, YbAg and YbAu. The various corrections to the baseline values are shown.}
\label{tab:finalresults}
\renewcommand*{\arraystretch}{1.3}
\begin{tabular}{l @{\hskip 25pt} r r r @{\hskip 25pt} r r r}
 \toprule
  & \multicolumn{3}{c}{\hspace{-30pt} $W_\mathrm{d} \, \bigl[10^{24}\frac{h\,\text{Hz}}{e\,\text{cm}}\bigr]$} & \multicolumn{3}{c}{$W_\mathrm{s} \, [h\,\text{kHz}]$}\\
  \cmidrule(lr{0.4in}){2-4}\cmidrule(lr{0.05in}){5-7}
  & YbCu & YbAg & YbAu & YbCu & YbAg & YbAu \\ 
 \midrule
 Baseline values & 13.122 & 11.869 & 1.326 & --47.647 & --44.361 & 6.979\\[1ex]
\textbf{Corrections}&  &  &  &  &  & \\
 Basis set (v4z vs v3z) & 0.020 & 0.035 & 0.193 & --0.269 & --0.238 & --0.983\\
 Active space (all-electron/$+6000\,E_h$ vs baseline) & 0.149 & 0.135 & 0.105 & --0.517 & --0.483 & 1.478\\
 Higher excitations (FSCCSDT vs FSCCSD) & --0.013 & 0.139 & 0.495 & --0.047 & --0.509 & --2.117\\
 Vibrational effects  & --0.032 & --0.024 & 0.013 & 0.115 & 0.086 & --0.048\\
 \midrule
 Recommended values & 13.245 & 12.154 & 2.131 & --48.365 & --45.505 & 5.309\\
 \bottomrule
\end{tabular}
\end{table*}

\begin{figure}[bt]
    \centering
    \includegraphics[width=.48\textwidth]{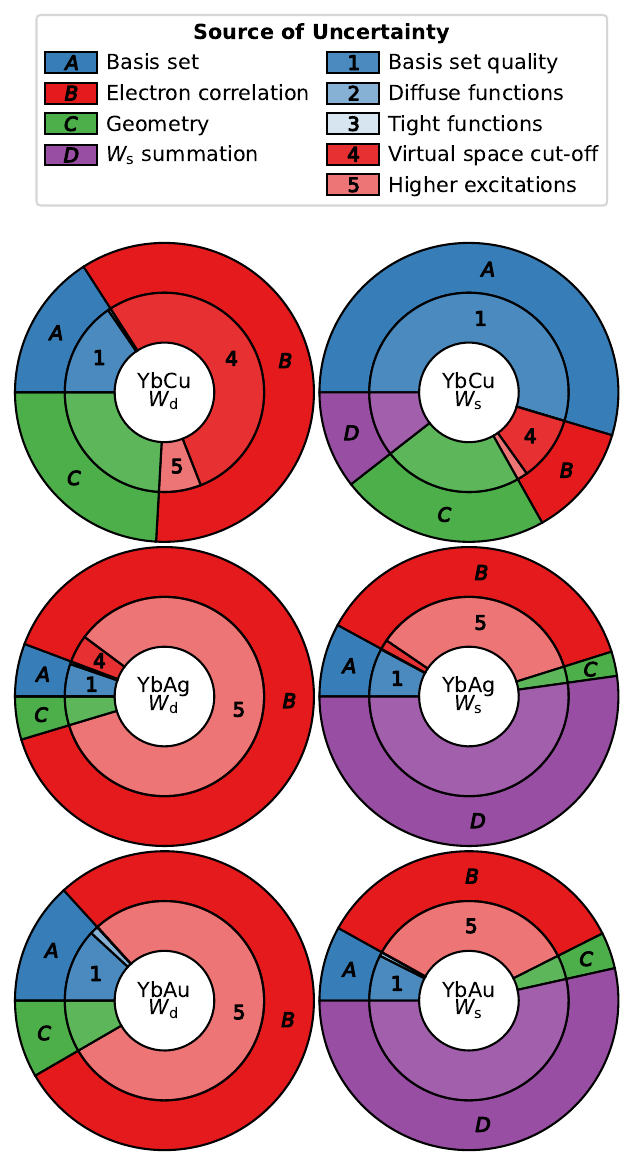}
    \caption{The distribution of all sources of uncertainty, given as a percentage of the total uncertainty of $W_\mathrm{d}$ and $W_\mathrm{s}$ for YbCu, YbAg and YbAu. The outer donut charts show the main uncertainty sources, while the inner donut charts show the individual sources of uncertainty.}
    \label{fig:pie_charts}
\end{figure}

To estimate conservative and reliable uncertainties for these values, a similar treatment to that given in Refs.~\citenum{haase_2021_systematic} and \citenum{chamorro_2022_molecular} will be employed. The individual uncertainties obtained from the considerations discussed in the previous sections are given in Table \ref{tab:uncertainty}, graphically displayed in Figure~\ref{fig:pie_charts}, and analyzed in the following subsections. 

\begin{table*}[bt]
\caption{The various sources of uncertainty for $W_\mathrm{d}$ and $W_\mathrm{s}$ of YbCu, YbAg and YbAu. The uncertainties are assumed to be independent, and the total uncertainty of the final results are given accordingly.}
\label{tab:uncertainty}
\begin{tabular}{@{\extracolsep{\fill}} l @{\hskip 25pt} *{3}{l} @{\hskip 25pt} *{3}{l}}
 \toprule
 &\multicolumn{3}{c}{\hspace{-30pt} $\delta W_\mathrm{d}\,\bigl[10^{24}\frac{h\,\text{Hz}}{e\,\text{cm}}\bigr]$}&\multicolumn{3}{c}{$\delta W_\mathrm{s}\,\bigl[h\,\text{kHz}\bigr]$}\\\cmidrule(lr{0.4in}){2-4}\cmidrule(lr{0.05in}){5-7}
 
 Uncertainty source& YbCu & YbAg & YbAu & YbCu & YbAg & YbAu\\
 \midrule
 \multicolumn{7}{l}{\textbf{Basis set}}\\[0.5ex]
 Basis set quality &0.010 & 0.018 & 0.097 & 0.135 & 0.120 & \phantom{0}0.492\\
 
 Diffuse functions &0.002 & 0.005 & 0.033 & 0.003 & 0.013 & \phantom{0}0.143 \\
 
 Tight functions &0.001 & 0.002 & 0.006 & 0.001 & 0.006 & \phantom{0}0.024 \\[0.5ex]

 \multicolumn{7}{l}{\textbf{Electron correlation}}\\[0.5ex]
 
 Virtual space cut-off  &0.019 & 0.016 & 0.005 & 0.059 & 0.054 & \phantom{0}0.003 \\
 
 Higher excitations &0.007 & 0.070 & 0.248 & 0.024 & 0.255 & \phantom{0}1.059 \\[1ex]

 \textbf{Geometry} &0.013 & 0.017 & 0.081 & 0.087 & 0.070 & \phantom{0}0.358 \\[1ex]

 \textbf{Sum of $W_\text{s}$ approx.} &-- & -- & -- & 0.059 & 0.308 & \phantom{0}1.313 \\
 \midrule
 \multicolumn{7}{l}{\textbf{Total uncertainty}}\\
 Absolute uncertainty &0.026 & 0.076 & 0.280 & 0.182 & 0.426 & \phantom{0}1.799 \\
 Relative uncertainty [\%] & 0.19 & 0.62 & 13.13 & 0.38 & 0.94 & 33.94 \\
 \bottomrule
\end{tabular}
\end{table*}

\subsubsection{Basis set}

The uncertainty due to the basis set has three different sources: (i) the general quality of the basis set, (ii) the quality of the treatment of tight functions, and (iii) the quality of the treatment of diffuse functions. 

The uncertainties in the general quality of the basis sets are due to the fact that the values possibly have not converged yet at the v4z level. The baseline values are corrected by adding the difference between the v4z and v3z results, as found in Section~\ref{sec:basisset}. For all parameters, the uncertainties in basis set quality are taken as a half of the differences between v4z and v3z results, to account for the unconsidered effects arising from the use of larger basis sets.

The uncertainties due to the possible insufficient amount of tight functions are given by the differences between the cv3z and the v3z results. Finally, the differences between the s-aug-v3z and v3z results determine the uncertainties in the quality of diffuse function treatment.

\subsubsection{Electron correlation}

Electron correlation is affected by the chosen (occupied and virtual) active space cut-offs, along with the excitations taken into account. 

To account for the frozen orbitals in the baseline calculation, the results provided in the Supplementary Material are used. The difference in enhancement factors between correlating all electrons, and freezing 2, 4 and 56 electrons for YbCu, YbAg and YbAu, respectively, was used to correct the baseline values.

The correction from the employed virtual space cut-off is given by the differences between the obtained values at a virtual space cut-off of $6000\,E_h$, and the values found using an active space cut-off of $\pm3000\,E_h$. To account for the effect of higher lying virtual orbitals, half of this difference is taken as an additional uncertainty.

The baseline values do not take into account CC triple excitations. The differences between the results obtained using the FSCCSDT and the FSCCSD methods, as given in Section~\ref{sec:methods}, were used to correct for this omission. Since these results may not be fully converged yet, the effect of including triple excitations is used to estimate the uncertainty arising from not including quadruple and higher excitations. The obtained differences between the FSCCSDT and FSCCSD results is multiplied one half to account for the uncertainty due to excitations involving more than three electrons.

\subsubsection{Geometry}

The bond length uncertainty contributes to the enhancement factor uncertainty due to the geometry of the systems. To estimate the uncertainty due to the bond length, the enhancement factors were computed for internuclear distances $\SI{0.01}{\angstrom}$ larger and smaller than the calculated $R_e$. The largest deviations from the parameters at equilibrium bond lengths were taken as their uncertainties.

The vibrational corrections obtained in Section~\ref{sec:vibcorr} are applied to the baseline values, as can be seen in Table \ref{tab:finalresults}.

\subsubsection{Sum of atomic $W_\text{s}$ approximation}

Due to the fact that the interaction constants $k_{\mathrm{s},K}$ in Eq.~\eqref{eq:HSPS-2} are specific to each nucleus, calculating the total $W_\text{s}$ value as a sum of the atomic contribution is only approximate and introduces an associated error. We analyze this error in detail in an upcoming publication \cite{Wserror}. Here, we provide the resulting uncertainties of 0.059, 0.308 and 1.318~$h$~kHz for YbCu, YbAg and YbAu, respectively. We note that the specific errors differ for each isotopologue. Therefore, here we use the isotopic average weighted by the natural abundances of all constituting elements.

\subsubsection{Total uncertainty}

To compute the total uncertainties for the reference values of the enhancement factors of YbCu, YbAg and YbAu, the Euclidean norm of the individual uncertainties is taken. These total uncertainties are obtained on the assumption that the different contributions are largely independent, since they concern high order effects. 

The relative uncertainties of $W_\mathrm{s}$ ($0.33\%$, $0.63\%$ and $22.74\%$, for YbCu, YbAg and YbAu, respectively) are larger than those found for $W_\mathrm{d}$ ($0.13\%$, $0.61\%$ and $13.12\%$, for YbCu, YbAg and YbAu, respectively).  For these systems, the enhancement factors are dominated by the uncertainty due to missing higher CC excitations and, to a smaller degree, by the basis set incompleteness, as illustrated in Figure \ref{fig:pie_charts}. 

The results for YbAu have significantly higher relative uncertainty than the other two systems. Both enhancement factors of this system are relatively small, due to a cancellation of similar sized contributions from the two constituent atoms, rendering them unstable and very sensitive to the computational settings and leading to a large relative uncertainty. The calculated $W_\mathrm{s}$ and $W_\mathrm{d}$ of YbCu and YbAg have remarkably small uncertainties of less than a single percent. For comparison, similar computational approach yielded uncertainties of 3 $-$ 7\% for the enhancement factors of BaF \cite{haase_2021_systematic}, YbCH$_3$ \cite{chamorro_2022_molecular}, and LuO \cite{chamorro2024paritytimereversalsymmetryviolation}. The lower uncertainty in the current case is partly due to the fact that we have corrected our results for the triple excitations, in contrast to the earlier works, where the missing excitations beyond doubles are a major source of uncertainty.

\subsection{Comparison to other systems}

The calculated enhancement factors are compared to those found for other ytterbium-containing molecules in Table \ref{tab:comparison}. Additionally, some of the systems currently used in experiments aiming to restrict the upper limit on the eEDM were added. The $W_\mathrm{d}$ and $W_\mathrm{s}$ factors of YbCu and YbAg are of similar magnitude to those found for YbOH, YbCH$_3$ and YbF. Both enhancement factors of YbAu are significantly smaller than those found for any other ytterbium-containing system.

Since both the interactions of the eEDMs with electromagnetic fields and the S-PS-ne neutral-current interactions may contribute to an eventual experimental detection of $\mathcal{P,T}$-violating effects in molecules, these two type of interactions should be decoupled from each other. This can be done by performing measurements on systems with different enhancement factor ratios \cite{gaul_2019_systematic}. For the systems studied in this work, and some other molecules currently and previously under investigation, these ratios can be calculated from the values of enhancement factors reported in Table \ref{tab:comparison}.

\begin{table}[bt]
\centering
\caption{\label{tab:comparison} Reference values of $W_\mathrm{d}$ and $W_\mathrm{s}$ for YbCu, YbAg and YbAu compared to other Ytterbium-containing molecules and systems currently or previously investigated to determine the lowest upper limits on $d_\mathrm{e}$ and $k_\mathrm{s}$.}
\renewcommand*{\arraystretch}{1.3}
\begin{threeparttable}
\begin{tabular*}{\linewidth}{@{\extracolsep{\fill}} l@{\hskip 1pt} l@{\hskip 1pt} l l}
 \toprule
 System & Source 
 & $ W_\mathrm{d} \, \bigl[10^{24}\frac{h\,\text{Hz}}{e\,\text{cm}}\bigr]$ 
 & $ W_\mathrm{s} \, [h\,\text{kHz}]$\\
 \midrule
 YbCu &This work & 13.24(3) & \phantom{0}--48.36(18)\\
 YbAg &This work & 12.15(8) & \phantom{0}--45.5(4)\\
 YbAu &This work & \phantom{0}2.13(28) & \phantom{--10}5.3(18)\\
 YbOH  & Ref.~\citenum{denis_2019_enhancement}           & 11.32(48) &               \\
 YbCH$_3$ & Ref.~\citenum{chamorro_2022_molecular}       & 13.80(35) & \phantom{0}--50.16(127) \\
 YbF      & Ref.~\citenum{nayak_2009_reappraisal}        & 11.64         &       \\
          & Ref.~\citenum{nayak_2007_iab}                &               & \phantom{0}--41.2  \\
          & Ref.~\citenum{Abe2014}                       & 11.17(89) &       \\
          & Ref.~\citenum{Sunaga2016}                    & 11.23          & \phantom{0}--40.52(324) \\
RaAg      & Ref.~\citenum{fleig_2021_theoretical}        & 30.9           & --175.1          \\
HfF$^+$  & Ref.~\citenum{fleig_2017_odd}                & 10.98          & \phantom{--1}20.0            \\
 ThO      & Ref.~\citenum{skripnikov_2013_communication} & 20             & \phantom{--}116             \\
 BaF      & Ref.~\citenum{haase_2021_systematic}         & \phantom{0}3.13(12)  & \phantom{--10}8.29(12)   \\
          & Ref.~\citenum{Talukdar_2020}                 & \phantom{0}3.15(30)  & \phantom{--10}8.35(70)   \\
          & Ref.~\citenum{nayak_2007_iab}                &               & \phantom{10}--9.7            \\
%
 \bottomrule
\end{tabular*}
\end{threeparttable}
\end{table}

\section{Conclusion}\label{sec:conclusion}

In order to extract information about the $\mathcal{P,T}$-violating effects arising from both the interactions between eEDMs and electromagnetic fields and the S-PS-ne neutral-current interactions from precision experiments in paramagnetic polar molecules containing only non-zero nuclear spins, the enhancement factors $W_\mathrm{d}$ and $W_\mathrm{s}$ have to be obtained through molecular electronic structure computations. In this work, these parameters were computed  for the YbCu, YbAg and YbAu systems, selected due to their possible experimental advantages. The enhancement factors $W_\mathrm{d}$ were calculated using two different schemes, and here we report (to the best of our knowledge) the first 4C computations using scheme 1. Besides, a thorough uncertainty analysis was performed to assign a conservative error on the obtained results.

The recommended values were calculated using the FSCC method and the 4C DC Hamiltonian in conjunction with relativistic basis sets. The main contributing sources of uncertainty are due to the limited basis set sizes and the neglect of CC excitations beyond triples.

The obtained enhancement factors of YbCu and YbAg are of very similar size to other Yb-containing compounds investigated in the literature. In case of YbAu, the cancellation of the contributions arising from the two nuclei in the system leads to vanishingly small total $W_\mathrm{d}$ and $W_\mathrm{s}$ values. For YbCu and YbAg, the results are also of similar size as for other systems currently investigated experimentally to search for signs of $\mathcal{P,T}$-violating effects. Compared to YbF and YbOH, the alternate method of producing and cooling these systems provides an alternative route for future experiments setting a lowest upper limit on the eEDM.

\section*{Supplementary Material}
In the Supplementary Material we analyze the influence of the active space (in particular, of the energy of occupied and virtual correlated orbitals) on the calculation of $W_\mathrm{d}$ and $W_\mathrm{s}$. We also present a set of tables where we report the values of these parameters using different nuclear models, different schemes corresponding to the use of the two effective Hamiltonians described in this work (for the case of $W_\mathrm{d}$), and a few different basis sets. In addition, tables are provided showing the dependence of the molecular enhancement factors on the use of different methods to treat electron correlation, on vibrational effects, and also the contributions to $W_\mathrm{d}$ and $W_\mathrm{s}$ associated with each nucleus of the studied molecular systems.

\section*{Acknowledgments}

We would like to thank the University of Groningen's Center for Information Technology and the Dutch National Supercomputer for their support and for providing access to the Hábrók and Snellius high-performance computing clusters. This work made use of the Dutch national e-infrastructure with the support of the SURF Cooperative using grants no.~EINF-5787, EINF-8014 and EINF-8532. IAA thanks R.~Berger and K.~Gaul for inspiring discussions, and acknowledges partial support from FONCYT through grants PICT-2021-I-A-0933 and PICT-2020-SerieA-0052, and CONICET through grant PIBAA-2022-0125CO. The work of AB, IAA and SH was supported by the project \textit{Probing Particle Physics with Polyatomic molecules} with project number OCENW.M.21.098 of the research programme M2 which is financed by the Dutch Research Council (NWO). The work of AB was supported by the project \textit{High Sector Fock space coupled cluster method: benchmark accuracy across the periodic table} with project number Vi.Vidi.192.088 of the research programme Vidi which is financed by the Dutch Research Council (NWO). The work of LFP and SH was supported by the project \textit{Searching for missing antimatter with trapped molecules} with project number VI.C.212.016 of the research programme Vici which is financed by the Dutch Research Council (NWO). LFP acknowledges the support from the Slovak Research and Development Agency projects APVV-20-0098 and APVV-20-0127.

\bibliography{Yb}

\end{document}


\title{Supplementary Material -- $\mathcal{P,T}$-odd effects in YbCu, YbAg and YbAu}

\author{Johan David Polet}
\affiliation{\RUG}

\author{Yuly Chamorro}
\affiliation{\RUG}
\affiliation{\Nikhef}

\author{Lukáš F. Pašteka}
\affiliation{\RUG}
\affiliation{\Nikhef}
\affiliation{\CU}

\author{Steven Hoekstra}
\affiliation{\RUG}
\affiliation{\Nikhef}

\author{Michał Tomza}
\affiliation{\UW}

\author{Anastasia Borschevsky}
\affiliation{\RUG}
\affiliation{\Nikhef}

\author{I.~Agustín Aucar}
\email[Author to whom correspondence should be addressed. Electronic mail: ]{agustin.aucar@conicet.gov.ar}
\affiliation{\RUG}
\affiliation{\Nikhef}
\affiliation{\IMIT}


\date{\today}

\maketitle
\tableofcontents

\subsection{Correlation active space}\label{sec:elec-corr}

To reduce computational effort and make these calculations tractable, we have reduced both the amount of correlated electrons and the size of the correlated virtual space. In the following, we examine the effect of these measures on the calculated enhancement factors.

\subsubsection{Correlated occupied orbitals}\label{sec:occupied}

\begin{figure}[H]
    \centering
    \includegraphics[width=.45\textwidth]{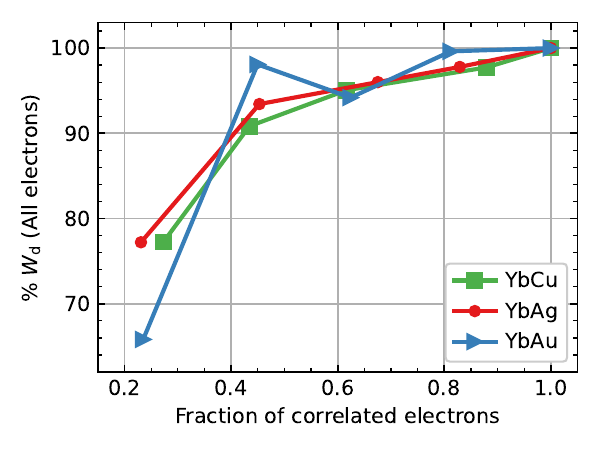}
    \caption{Graphical representation of the behavior of $W_\mathrm{d}$ of YbCu, YbAg and YbAu for varying numbers of correlated electrons. The enhancement factor is given as a percentage of the result obtained when correlating all electrons. These results were obtained on the FSCCSD/v2z level with a symmetric virtual space cut-off.}
    \label{fig:active_space_cut-off_Wd}
\end{figure}
\begin{figure}[H]
    \centering
    \includegraphics[width=.45\textwidth]{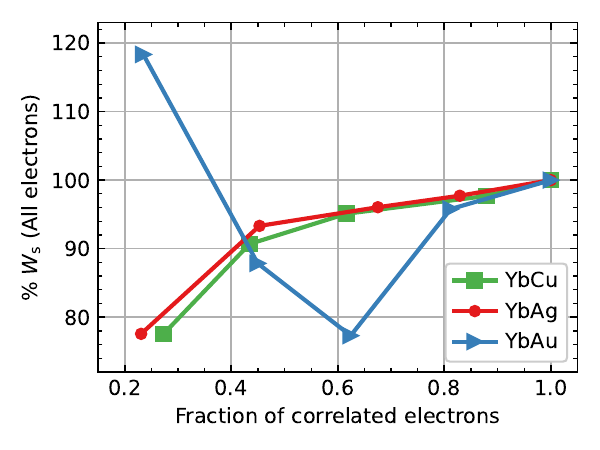}
    \caption{Graphical representation of the behavior of $W_\mathrm{s}$ of YbCu, YbAg and YbAu for varying numbers of correlated electrons. The enhancement factor is given as a percentage of the result obtained when correlating all electrons. These results were obtained on the FSCCSD/v2z level with a symmetric virtual space cut-off.}
    \label{fig:active_space_cut-off_Ws}
\end{figure}

Table \ref{tab:freezeCu}, Figure~\ref{fig:active_space_cut-off_Wd}, and Figure~\ref{fig:active_space_cut-off_Ws} show the behavior of the enhancement factors with increasing number of correlated electrons. The virtual space cut-off --which dictates the number of virtual orbitals to include-- was set to be symmetric in terms of energy with the active space cut-off. The results were obtained using the FSCCSD method and the v2z basis set.

When halving the number of correlated electrons, all enhancement factors remain within 23\% of the results found when correlating all electrons. Freezing further electrons has a more significant effect on the parameters. The behavior of both parameters is similar for YbCu and YbAg, where the contribution from the secondary nucleus is small, as can be observed in Figure~\ref{fig:active_space_cut-off_Wd} and Figure~\ref{fig:active_space_cut-off_Ws}. Because of the cancellation of contributions, the trend is more chaotic for YbAu, in particular for $W_\mathrm{s}$.

\begin{center}
\begin{table*}[htp]
\renewcommand*{\arraystretch}{1.3}
\caption{The enhancement factors of YbCu, YbAg and YbAu computed with different numbers of correlated electrons, $N_{\textit{el}}^{\textit{co}}$. The factors were computed at the FSCCSD/v2z level.}
\label{tab:freezeCu}
\centering
\begin{tabular}{l @{\hskip 15pt} l @{\hskip 5pt} l @{\hskip 15pt} l @{\hskip 15pt} r @{\hskip 5pt} r @{\hskip 15pt} r @{\hskip 5pt} r}
\toprule
  $N_{\textit{el}}^{\textit{co}}$ & \multicolumn{2}{l}{Frozen orbitals\;\;} & Cut-offs&\hskip 30pt$W_\mathrm{d}$& &\hskip 30pt$W_\mathrm{s}$ &\\\cmidrule(lr{0.2in}){2-3}\cmidrule(lr{0.2in}){5-6}\cmidrule(lr){7-8}
 
  & Yb & Cu &  [$E_h$]&$\bigl[10^{24}\frac{h\,\text{Hz}}{e\,\text{cm}}\bigr]$ & [\%]&$\bigl[h\,\text{kHz}\bigr]$ & [\%]\\
 \midrule
 27& [Xe]          & [Ar]          & \phantom{000}1 & \phantom{0}9.890 & --22.75 & --34.443 & --22.46\\
 43& [Kr]4d$^{10}$ & [Ne]          & \phantom{000}6 & 11.629           & \phantom{0}--9.16 & --40.278 & \phantom{0}--9.32\\
 61& [Ar]3d$^{10}$ & [Ne]          & \phantom{00}20 & 12.169           & \phantom{0}--4.95 & --42.239 & \phantom{0}--4.91\\
 87& [Ne]          & [He]          & \phantom{0}100 & 12.512           & \phantom{0}--2.27 & --43.390 & \phantom{0}--2.31\\
 97& [He]          & \phantom{[]}-- & \phantom{0}500 & 12.691           & \phantom{0}--0.87 & --44.019 & \phantom{0}--0.90\\
 99& \phantom{[]}-- & \phantom{[]}-- & 3000 & 12.803           & \phantom{[]}--- & --44.418 & \phantom{[]}--- \\
 \midrule
 $N_{\textit{el}}^{\textit{co}}$ & Yb & Ag & & & \\
 \midrule
 27&[Xe] & [Kr] & \phantom{000}0.8 & \phantom{0}9.063 &--22.78 & --32.395 & --22.43\\
 53&[Kr] & [Ar]3d$^{10}$ & \phantom{00}10 & 10.966 &\phantom{0}--6.57 & --38.974 & \phantom{0}--6.68\\
 79&[Ar]3d$^{10}$ & [Ne] & \phantom{00}30 & 11.268 &\phantom{0}--3.99 & --40.112 & \phantom{0}--3.95\\
 97&[Ne] & [Ne] & \phantom{0}100 & 11.477 &\phantom{0}--2.21 & --40.813 & \phantom{0}--2.28\\
 113& [He] & [He] & \phantom{0}500 & 11.634 & \phantom{0}--0.87 & --41.387 & \phantom{0}--0.90\\
 117&\phantom{[]}-- & \phantom{[]}-- & 3000 & 11.737 & \phantom{[]}--- & --41.763 & \phantom{[]}---\\
 \midrule
$N_{\textit{el}}^{\textit{co}}$ & Yb & Au & & & \\
 \midrule
 35&[Kr]3d$^{10}$ & [Xe]3f$^{14}$ & \phantom{000}2.68 & 1.071 &--34.18 & 7.683 & \phantom{--}18.34\\
 67&[Kr] & [Kr]4d$^{10}$ & \phantom{00}10 & 1.595 &\phantom{0}--1.91 & 5.704 & --12.15\\
 93&[Ar]3d$^{10}$ & [Ar]3d$^{10}$ & \phantom{00}40 & 1.532 &\phantom{0}--5.81 & 5.019 & --22.69\\
 121&[Ne] & [Ne] & \phantom{00}95 & 1.620 &\phantom{0}--0.38 & 6.219 & \phantom{0}--4.22\\
 149&\phantom{[]}-- & \phantom{[]}-- & 3000 & 1.626 & \phantom{[]}--- & 6.493 & \phantom{[]}---\\
 \bottomrule
\end{tabular}
\end{table*}
\end{center}

The final results will be computed with 2, 4 and 56 electrons frozen for YbCu, YbAg and YbAu, respectively. For YbCu and YbAg, both parameters are only underestimated by $\sim0.9\%$ at this level. For YbAu, correlating more electrons with a larger basis set is currently not computationally possible due to the large size of these systems. Any under-representation will be dealt with by correcting the baseline values.


\subsubsection{Correlated virtual orbitals}

\begin{figure}[H]
    \centering
    \includegraphics[width=.45\textwidth]{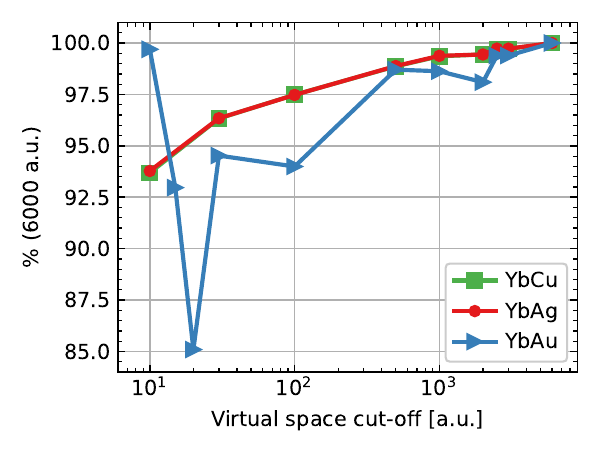}
    \caption{The $W_\mathrm{d}$ enhancement factor of YbCu, YbAg and YbAu, for various virtual space cut-offs, compared to the result with a virtual space cut-off at 6000~$E_h$. Computations were performed correlating all electrons at the FSCCSD/v2z level.}
    \label{fig:virtual_space_cut-off}
\end{figure}

Including more virtual orbitals has a noticeable impact on the computed enhancement factors. Figure~\ref{fig:virtual_space_cut-off} shows the behavior of $W_\mathrm{d}$ for increasing virtual space cut-off. 
The $W_\mathrm{d}$ results are given as a percentage of the result found with a virtual space cut-off at 6000~$E_h$. It is assumed that at this virtual space cut-off, the enhancement factor has converged. The computations were performed with all electrons correlated on the FSCCSD/v2z level. It should be noted that lower lying virtual orbitals lie closer together than higher lying virtual orbitals. For YbCu, for example, only 14 orbitals are present between the 3000~$E_h$ cut-off and the 6000~$E_h$ cut-off, while between 10~$E_h$ and 30~$E_h$ there are 62 orbitals. 

Again, the behavior for YbCu and YbAg is similar, with $W_\mathrm{d}$ decreasing when reducing the number of included virtual orbitals. Only when the virtual orbital cut-off is set below 500~$E_h$ does $W_\mathrm{d}$ drop below $\sim1\%$ of its value at 6000~$E_h$ for both systems. The behavior of YbAu is again significantly different because of the partial cancellation of individual contributions.

The results for $W_\mathrm{s}$ with a virtual orbital cut-off of 3000~$E_h$ and 6000~$E_h$ are comparable in percentage to those found for $W_\mathrm{d}$, as can be seen in Table \ref{tab:virtual}. Since these are the only relevant values for the uncertainty determination, computations with lower virtual space cut-offs were not performed for $W_\mathrm{s}$. 

For the baseline computations, symmetric virtual orbital cut-offs were selected. These were $\pm500$~$E_h$, $\pm500$~$E_h$ and $\pm40$~$E_h$ for YbCu, YbAg and YbAu, respectively. The effect of excluded higher lying virtual orbitals will be accounted for in the calculation of the final values.

\begin{center}
\begin{table}[H]
\renewcommand*{\arraystretch}{1.3}
\caption{The enhancement factors of YbCu, YbAg and YbAu computed with different virtual orbital cut-offs. The factors were computed at the FSCCSD/v2z level, correlating all the electrons.}
\label{tab:virtual}
\centering
\begin{tabular}{l @{\hskip 10pt} l @{\hskip 15pt} ll @{\hskip 15pt} ll}
\toprule
 & & \phantom{0}$W_\mathrm{d}$ & & \phantom{0}$W_\mathrm{s}$ & \\
 \cmidrule(lr{0.2in}){3-4}\cmidrule(lr){5-6}
 System & Cut-off [$E_h$] &$\bigl[10^{24}\frac{h\,\text{Hz}}{e\,\text{cm}}\bigr]$ & [\%]& $\bigl[h\,\text{kHz}\bigr]$ & [\%]\\
 \midrule
 YbCu & 3000 & 12.803 & --0.29 & --44.418 & --0.26\\
      & 6000 & 12.840 & \phantom{[]}--- & --44.539 & \phantom{[]}---\\
 \midrule
 YbAg & 3000 & 11.737 & --0.27 & --41.763 & --0.26\\
      & 6000 & 11.769 & \phantom{[]}--- & --41.870 & \phantom{[]}---\\
 \midrule
 YbAu & 3000 & \phantom{0}1.626 & --0.61 & \phantom{--0}6.493 & --0.08\\
      & 6000 & \phantom{0}1.636 & \phantom{[]}--- & \phantom{--0}6.498 & \phantom{[]}---\\
 \bottomrule
\end{tabular}
\end{table}
\end{center}

\section{Additional Tables}

\begin{table}[H]
\centering
\caption{Tabular representation of the behavior of $W_\mathrm{d}$ for YbCu, YbAg and YbAu using Gaussian-type and point-type nuclear models (GN and PN, respectively). These results were obtained on the FSCCSD/v2z level, correlating all the electrons, and with a virtual space cut-off at 3000~$E_h$.}
\label{tab:NSeff}
\renewcommand*{\arraystretch}{1.3}
\begin{tabular}{l@{\hskip 25pt} l @{\hskip 20pt} l}
 \toprule
 & \multicolumn{2}{c}{$W_\mathrm{d}\,\bigl[10^{24}\frac{h\,\text{Hz}}{e\,\text{cm}}\bigr]$}\\\cmidrule(lr){2-3}
 Molecule & PN & GN\\
 \midrule
 YbCu & 12.939 & 12.803\\
 YbAg & 11.870 & 11.737\\
 YbAu & \phantom{0}1.488 & \phantom{0}1.626\\
 \bottomrule
\end{tabular}
\end{table}

\begin{table*}[htp]
\centering
\caption{Contributions to $W_\mathrm{s}$ for YbCu, YbAg and YbAu, arising from each individual nucleus. The results were obtained on the FSCCSD/v2z level of theory with all electrons correlated and a virtual space cut-off of 3000~$E_h$. A Gaussian-type and a point-like nuclear models (GN and PN, respectively) were used. The sum of individual contributions $W_\mathrm{{s},\text{Yb}}+W_{\mathrm{s},X}$ ($X=$ Cu, Ag, Au) is also given.}
\label{tab:Ws-indiv}
\renewcommand*{\arraystretch}{1.3}
\begin{tabular}{l@{\hskip 25pt} l @{\hskip 20pt} l @{\hskip 20pt} l @{\hskip 20pt} l @{\hskip 20pt} l @{\hskip 20pt} l}
 \toprule
 & \multicolumn{2}{c}{$W_\mathrm{{s},\text{Yb}}\,\bigl[10^{24}\frac{h\,\text{Hz}}{e\,\text{cm}}\bigr]$} & \multicolumn{2}{c}{$W_{\mathrm{s},X}\,\bigl[10^{24}\frac{h\,\text{Hz}}{e\,\text{cm}}\bigr]$} & \multicolumn{2}{c}{$W_\mathrm{{s},\text{Yb}}+W_{\mathrm{s},X}\,\bigl[10^{24}\frac{h\,\text{Hz}}{e\,\text{cm}}\bigr]$}\\\cmidrule(lr){2-3}\cmidrule(lr){4-5}\cmidrule(lr){6-7}
 Molecule & PN & GN & PN & GN & PN & GN\\
 \midrule
 YbCu & --51.563 & --44.656 & \phantom{0}0.238 & \phantom{0}0.238 & --51.325 & --44.418\\
 YbAg & --50.690 & --43.900 & \phantom{0}2.217 & \phantom{0}2.137 & --48.473 & --41.763\\
 YbAu & --51.966 & --45.002 & 64.094 & 51.495 & \phantom{--}12.128 & \phantom{--0}6.493\\
 \bottomrule
\end{tabular}
\end{table*}

\begin{table}[H]
\centering
\caption{Tabular representation of the behavior of $W_\mathrm{d}$ of YbCu, YbAg and YbAu computed using scheme 1 (S1) and scheme 2 (S2). For S1, the contributions from both nuclei are shown (Yb and $X=$ Cu, Ag, Au). These results were obtained on the FSCCSD/v3z level, with symmetric space cut-offs of $\pm$500~$E_h$ for YbCu and YbAg, and $\pm$40~$E_h$ for YbAu. Point-like nuclear models were used in all cases.}
\label{tab:stratagems}
\renewcommand*{\arraystretch}{1.3}
\begin{tabular}{l@{\hskip 25pt} l @{\hskip 20pt} l @{\hskip 20pt} l @{\hskip 20pt} l}
 \toprule
 & \multicolumn{4}{c}{$W_\mathrm{d}\,\bigl[10^{24}\frac{h\,\text{Hz}}{e\,\text{cm}}\bigr]$}\\\cmidrule(lr){2-5}
 Molecule & S1 (Yb) & S1 ($X$) & S1 (Total) & S2\\
 \midrule
 YbCu & 13.807 & \phantom{0}--0.166 & 13.640 & 13.435\\
 YbAg & 13.489 & \phantom{0}--1.141 & 12.348 & 12.172\\
 YbAu & 13.330 & --12.143 & \phantom{0}1.188 & \phantom{0}1.131\\
 \bottomrule
\end{tabular}
\end{table}

\begin{center}
\begin{table*}[htp]
\caption{{Enhancement factors of YbCu, YbAg and YbAu, for v2z, v3z, v4z, cv3z and s-aug-v3z basis sets. Computations were performed at the FSCCSD level, with active space cut-offs set to $\pm$20~$E_h$ for YbCu and $\pm$10~$E_h$ for YbAg and YbAu. For $W_\mathrm{s}$, the individual contributions and the sum $W_\mathrm{{s},\text{Yb}}+W_{\mathrm{s},X}$ ($X=$ Cu, Ag, Au) are given.}}
\label{tab:basis}
\renewcommand*{\arraystretch}{1.3}
\centering
\begin{tabular}{l @{\hskip 10pt} lll @{\hskip 10pt} lll @{\hskip 10pt} lll @{\hskip 10pt} lll} 
 \toprule
  & \multicolumn{3}{c}{$W_\mathrm{d}\,\bigl[10^{24}\frac{h\,\text{Hz}}{e\,\text{cm}}\bigr]$} & \multicolumn{3}{c}{\;\;$W_\mathrm{{s},\text{Yb}}\,[h\,\text{kHz}]\;\;$} & \multicolumn{3}{c}{\;\;$W_{\mathrm{s},X}\,[h\,\text{kHz}]\;\;$} & \multicolumn{3}{c}{\;\;$W_\mathrm{{s},\text{Yb}}+W_{\mathrm{s},X}\,[h\,\text{kHz}]\;\;$} \\\cmidrule(lr{0.2in}){2-4}\cmidrule(lr{0.2in}){5-7}\cmidrule(lr{0.2in}){8-10}\cmidrule(lr){11-13}
 \rule{0pt}{2.5ex}Method & YbCu & YbAg & YbAu & \;YbCu & YbAg & YbAu & YbCu & YbAg & YbAu & \;YbCu & YbAg & YbAu \\[0.5ex]
 \midrule
 v2z & 12.169 & 10.966 & 1.595 & --42.457 & --40.930 & --41.780 & 0.218 & 1.955 & 47.484 & --42.239 & --38.974 & 5.704\\
 
 v3z & 12.594 & 11.158 & 1.122 & --46.011 & --44.041 & --44.770 & 0.256 & 2.341 & 52.264 & --45.755 & --41.700 & 7.494\\
 
 v4z & 12.614 & 11.193 & 1.314 & --46.292 & --44.398 & --45.076 & 0.268 & 2.459 & 51.587 & --46.024 & --41.939 & 6.511\\

 \midrule

 cv3z & 12.594 & 11.159 & 1.127 & --46.011 & --44.048 & --44.763 & 0.256 & 2.342 & 52.233 & --45.755 & --41.706 & 7.470\\

 s-aug-v3z & 12.593 & 11.153 & 1.089 & --46.010 & --44.033 & --44.735 & 0.257 & 2.345 & 52.371 & --45.753 & --41.688 & 7.637\\
 \bottomrule
\end{tabular}
\end{table*}
\end{center}

\begin{figure}[H]
    \centering
    \includegraphics[width=.45\textwidth]{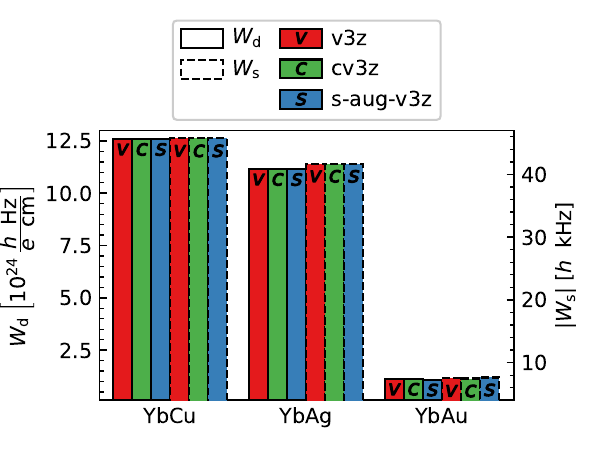}
    \caption{The enhancement factors of YbCu, YbAg and YbAu, using v3z, cv3z and s-aug-v3z basis sets. Computations were performed at the FSCCSD level with active space cut-offs set to $\pm$20~$E_h$ for YbCu and $\pm$10~$E_h$ for YbAg and YbAu.}
    \label{fig:basis_set_extra_functions}
\end{figure}

\begin{center}
\begin{table*}[htp]
\caption{Enhancement factors $W_\mathrm{d}$ and $W_\mathrm{s}$ of YbCu, YbAg and YbAu computed at the HF, MP2, FSCC-SD and FSCC-SDT levels of approach. Computations were done at v2z level, and the active space cut-off was set to $\pm$2~$E_h$ for the three systems. For $W_\mathrm{s}$, the individual contributions and the sums $W_\mathrm{{s},\text{Yb}}+W_{\mathrm{s},X}$ ($X=$ Cu, Ag, Au) are given.}
\label{tab:methods}
\renewcommand*{\arraystretch}{1.3}
\centering
\begin{tabular}{l @{\hskip 10pt} lll @{\hskip 10pt} lll @{\hskip 10pt} lll @{\hskip 10pt} lll} 
 \toprule
  & \multicolumn{3}{c}{$W_\mathrm{d}\,\bigl[10^{24}\frac{h\,\text{Hz}}{e\,\text{cm}}\bigr]$} & \multicolumn{3}{c}{\;\;$W_\mathrm{{s},\text{Yb}}\,[h\,\text{kHz}]\;\;$} & \multicolumn{3}{c}{\;\;$W_{\mathrm{s},X}\,[h\,\text{kHz}]\;\;$} & \multicolumn{3}{c}{\;\;$W_\mathrm{{s},\text{Yb}}+W_{\mathrm{s},X}\,[h\,\text{kHz}]\;\;$} \\\cmidrule(lr{0.2in}){2-4}\cmidrule(lr{0.2in}){5-7}\cmidrule(lr{0.2in}){8-10}\cmidrule(lr){11-13}
 \rule{0pt}{2.5ex}Method & YbCu & YbAg & YbAu & YbCu & YbAg & YbAu & YbCu & YbAg & YbAu & YbCu & YbAg & YbAu \\[0.5ex]
 \midrule
 HF & \phantom{0}9.652 & \phantom{0}7.044 & \phantom{0}3.770 & --33.118 & --32.900 & --34.894 & 0.037 & 0.353 & 12.619 & --33.081 & --32.547 & --22.275\\
 
 MP2 & 11.174 & \phantom{0}9.188 & \phantom{0}2.314 & --38.745 & --38.401 & --38.674 & 0.075 & 0.705 & 22.852 & --38.670 & --37.696 & --15.822\\
 
 
 
 
 
 FSCCSD & 11.323 & 10.415 & \phantom{0}1.072 & --39.365 & --38.696 & --40.187 & 0.203 & 1.798 & 47.572 & --39.162 & --36.897 & \phantom{--0}7.385\\
 
 FSCCSDT & 11.310 & 10.555 & \phantom{0}1.567 & --39.435 & --39.167 & --40.710 & 0.225 & 1.761 & 45.978 & --39.210 & --37.406 & \phantom{--0}5.268\\
 \bottomrule
\end{tabular}
\end{table*}
\end{center}

\begin{center}
\begin{table*}[htp]
\caption{Enhancement factors $W_\mathrm{d}$ and $W_\mathrm{s}$ of YbCu, YbAg and YbAu, computed at different displacements ($\delta_R$) with respect to the calculated equilibrium bond lengths ($R_{e,\text{YbCu}}=2.7543~\unit{\angstrom}$, $R_{e,\text{YbAg}}=2.8589~\unit{\angstrom}$, and $R_{e,\text{YbAu}}= 2.6524~\unit{\angstrom}$). 
Computations were performed on the FSCCSD/v2z level of approach, with active space cut-offs set to $\pm$100~$E_h$ for YbCu and YbAg, and $\pm$95$E_h$ for YbAu.}
\label{tab:runcertainty}
\renewcommand*{\arraystretch}{1.3}
\centering
\begin{tabular}{l @{\hskip 20pt} lll @{\hskip 15pt} lll @{\hskip 15pt} lll} 
 \toprule
  & \multicolumn{3}{c}{\;\;$W_\mathrm{{s},\text{Yb}}\,[h\,\text{kHz}]\;\;$} & \multicolumn{3}{c}{\;\;$W_{\mathrm{s},X}\,[h\,\text{kHz}]\;\;$} & \multicolumn{3}{c}{\;\;$W_\mathrm{{s},\text{Yb}}+W_{\mathrm{s},X}\,[h\,\text{kHz}]\;\;$} \\\cmidrule(lr{0.2in}){2-4}\cmidrule(lr{0.2in}){5-7}\cmidrule(lr){8-10}
 \rule{0pt}{2.5ex}$\delta_R$ [\unit{\angstrom}] & YbCu & YbAg & YbAu & YbCu & YbAg & YbAu & YbCu & YbAg & YbAu \\[0.5ex]
 \midrule
 --0.10           & --44.447 & --43.714 & --44.309 & 0.262 & 2.290 & 54.502 & --44.186 & --41.423 & 10.193\\
 --0.05           & --44.051 & --43.317 & --44.145 & 0.247 & 2.178 & 52.251 & --43.805 & --41.138 & \phantom{0}8.106\\
 --0.01           & --43.711 & --42.974 & --43.998 & 0.235 & 2.093 & 50.579 & --43.476 & --40.881 & \phantom{0}6.581\\
 \phantom{--}0.00 & --43.623 & --42.885 & --43.959 & 0.232 & 2.073 & 50.178 & --43.390 & --40.813 & \phantom{0}6.219\\
 \phantom{--}0.01 & --43.533 & --42.795 & --43.919 & 0.230 & 2.052 & 49.785 & --43.303 & --40.742 & \phantom{0}5.866\\
 \phantom{--}0.05 & --43.159 & --42.418 & --43.749 & 0.219 & 1.973 & 48.282 & --42.940 & --40.445 & \phantom{0}4.532\\
 \phantom{--}0.10 & --42.657 & --41.914 & --43.513 & 0.206 & 1.879 & 46.553 & --42.451 & --40.036 & \phantom{0}3.041\\
\bottomrule
\end{tabular}
\end{table*}
\end{center}

\begin{table*}[htp]
\centering
\caption{Final recommended values of $W_\mathrm{{s},\text{Yb}}$ and $W_{\mathrm{s},X}$ for YbCu, YbAg and YbAu ($X=$ Cu, Ag, Au). The various corrections to the baseline values are shown.}
\label{tab:finalresults}
\renewcommand*{\arraystretch}{1.3}
\begin{tabular}{l @{\hskip 25pt} r r r @{\hskip 25pt} r r r}
 \toprule
  & \multicolumn{3}{c}{\hspace{-30pt} $W_\mathrm{{s},\text{Yb}} \, [h\,\text{kHz}]$} & \multicolumn{3}{c}{$W_{\mathrm{s},X} \, [h\,\text{kHz}]$}\\
  \cmidrule(lr{0.4in}){2-4}\cmidrule(lr{0.05in}){5-7}
  & YbCu & YbAg & YbAu & YbCu & YbAg & YbAu \\ 
 \midrule
 Baseline values & --47.920 & --46.845 & --46.300 & 0.273 & 2.484 & 53.278\\[1ex]
 \textbf{Corrections}&  &  &  &  &  & \\
 Basis set (v4z vs v3z) & --0.281 & --0.357 & --0.307 & 0.013 & 0.118 & --0.677 \\
 Active space (all-electron/$+6000~E_h$ vs baseline) & --0.521 & --0.511 & --1.912 & 0.004 & 0.028 & 3.391\\
 Higher excitations (FSCCSDT vs FSCCSD) & --0.070 & --0.472 & --0.523 & 0.022 & --0.038 & --1.594\\
 Vibrational effects & 0.118 & 0.096 & 0.029 & --0.002 & --0.010 & --0.077\\
 \midrule
 Recommended values & --48.674 & --48.089 & --49.013 & 0.310 & 2.582 & 54.321\\
 \bottomrule
\end{tabular}
\end{table*}

\begin{table*}[!htb]
\caption{The various sources of uncertainty for $W_\mathrm{{s},\text{Yb}}$ and $W_{\mathrm{s},X}$ of YbCu, YbAg and YbAu ($X=$ Cu, Ag, Au). The uncertainties are assumed to be independent, and the total uncertainty of the final results are given accordingly.}
\label{tab:uncertainty}
\begin{tabular}{@{\extracolsep{\fill}} l @{\hskip 25pt} *{3}{l} @{\hskip 25pt} *{3}{l}}
 \toprule
 &\multicolumn{3}{c}{\hspace{-30pt} $\delta W_\mathrm{{s},\text{Yb}}\,\bigl[h\,\text{kHz}\bigr]$}&\multicolumn{3}{c}{$\delta W_{\mathrm{s},X}\,\bigl[h\,\text{kHz}\bigr]$}\\\cmidrule(lr{0.4in}){2-4}\cmidrule(lr{0.05in}){5-7}
 
 Uncertainty source& YbCu & YbAg & YbAu & YbCu & YbAg & YbAu\\
 \midrule 
 \multicolumn{7}{l}{\textbf{Basis set}}\\[0.5ex]
 Basis set quality & 0.141 & 0.179 & 0.154 & 0.007 & 0.059 & 0.339\\
 
 Diffuse functions & 0.001 & 0.009 & 0.035 & 0.002 & 0.004 & 0.108\\
 
 Tight functions & 0.001 & 0.007 & 0.007 & 0.000 & 0.001 & 0.031\\[0.5ex]
 
 \multicolumn{7}{l}{\textbf{Electron correlation}}\\[0.5ex]
 
 Virtual space cut-off & 0.059 & 0.058 & 0.059 & 0.001 & 0.005 & 0.062\\
 
 Higher excitations & 0.035 & 0.236 & 0.262 & 0.011 & 0.019 & 0.798\\[1ex]
 \textbf{Geometry} & 0.090 & 0.090 & 0.040 & 0.003 & 0.021 & 0.397\\[1ex]
 \midrule
 \multicolumn{7}{l}{\textbf{Total uncertainty}}\\
 Absolute uncertainty & 0.180 & 0.314 & 0.313 & 0.013 & 0.065 & 0.961\\
 Relative uncertainty [\%] & 0.37 & 0.65 & 0.64 & 4.13 & 2.53 & 1.77\\
 \bottomrule
\end{tabular}
\end{table*}